\pdfoutput=1
\documentclass[cernpreprint, atlasdraft=false, texlive=2016, UKenglish, texmf]{atlasdoc}
 
\usepackage[subcaption]{atlaspackage}
\usepackage{atlasbiblatex}
 
\usepackage{atlasphysics}
 
\graphicspath{{logos/}{figures/}}
 
\usepackage{ANA-EXOT-2018-46-PAPER-defs}
 
\usepackage{multirow}
\usepackage{booktabs}
\usepackage{adjustbox}
\usepackage{makecell}
\usepackage{siunitx}
 
\usepackage{longtable}
 
\usepackage{enumitem}
 
\usepackage{xfrac}
 
\usepackage{diagbox}
\usepackage{gensymb}
\usepackage{cleveref}
\DeclareUnicodeCharacter{2009}{FIXME}
\usepackage{mathrsfs}
\usepackage{tabu}

\AtlasTitle{Search for dark matter produced in association with a Standard Model Higgs boson decaying into $b$-quarks using the full Run~2 dataset from the ATLAS detector}

\AtlasJournal{JHEP}
 
\AtlasJournalRef{JHEP 11 (2021) 209}
 
\AtlasDOI{10.1007/JHEP11(2021)209}
 
\AtlasAbstract{
The production of dark matter in association with Higgs bosons is predicted in several extensions of the Standard Model. An exploration of such scenarios is presented, considering final states with missing transverse momentum and $b$-tagged jets consistent with a Higgs boson. The analysis uses proton--proton collision data at a centre-of-mass energy of 13~TeV recorded by the ATLAS experiment at the LHC during Run~2, amounting to an integrated luminosity of 139~fb$^{-1}$. The analysis, when compared with previous searches, benefits from a larger dataset, but also has further improvements providing sensitivity to a wider spectrum of signal scenarios. These improvements include both an optimised event selection and advances in the object identification, such as the use of the likelihood-based significance of the missing transverse momentum and variable-radius track-jets. No significant deviation from Standard Model expectations is observed.
Limits are set, at 95\% confidence level, in two benchmark models with two Higgs doublets extended by either a heavy vector boson $Z'$ or a pseudoscalar singlet $a$ and which both provide a dark matter candidate $\chi$. In the case of the two-Higgs-doublet model with an additional vector boson $Z'$, the observed limits extend up to a $Z'$ mass of 3~TeV for a mass of 100~GeV for the dark matter candidate.
The two-Higgs-doublet model with  a dark matter particle mass of 10 GeV and an additional pseudoscalar $a$ is excluded for masses of the $a$
up to 520 GeV and 240 GeV for $\tan \beta = 1$ and $\tan \beta = 10$ respectively. Limits on the visible cross-sections are set and range from to 0.05~fb to 3.26~fb, depending on the missing transverse momentum and $b$-quark jet multiplicity requirements.
}
 
\author{The ATLAS Collaboration}
 
\AtlasRefCode{EXOT-2018-46}

\PreprintIdNumber{CERN-EP-2021-074}

\hypersetup{pdftitle={ATLAS document},pdfauthor={The ATLAS Collaboration}}
 
\begin{document}
 
\maketitle

\section{Introduction}
\label{sec:intro}
 
Various astrophysical observations based on gravitational interactions~\cite{Clowe:2003tk,Ade:2015xua} strongly support the existence of \ac{DM} which interacts through neither the strong nor the electromagnetic force.
However, the \ac{SM} provides no suitable \ac{DM} candidate particle.
There are many complementary search strategies for \ac{DM} including direct-detection~\cite{PhysRevLett.121.111302,PhysRevLett.117.121303,PhysRevLett.121.081307,PhysRevLett.123.241803,PhysRevD.100.022001} and indirect-detection experiments~\cite{Ackermann_2015} as well as searches at particle colliders.
 
Since \ac{DM} particles do not interact electromagnetically or strongly even direct-detection experiments have low efficiencies.
Therefore, instead of attempting to detect them directly, any \ac{DM} particles produced in proton--proton collisions at the \ac{LHC} would be deduced from an imbalance in the transverse momentum measured in that collision event (\MET).
This means that DM particles can only be detected if they are produced in association with visible particles.
When there is only one such particle, this gives rise to event topologies referred to as `mono-$X$' final states, where $X$ refers to the visible particle.
Prominent examples of these topologies are the mono-jet~\cite{Aad:2021egl,CMS-EXO-16-048,Aaltonen_2012}, mono-$Z/W$~\cite{EXOT-2016-23,CMS-EXO-16-037,HIGG-2016-28} and mono-photon~\cite{EXOT-2018-63,CMS-EXO-16-039,PhysRevLett.101.181602} final states.
 
Mono-$X$ topologies are typically dominated by cases where the visible particle is produced from \ac{ISR}, as the couplings between the \ac{SM} particle $X$ and the initial-state quarks or gluons are much larger than its couplings to the final-state \ac{DM} particles.
A counterexample is the case where the visible particle is a \ac{SM} Higgs boson (named the mono-Higgs signature).
Given that the coupling of the Higgs boson to light quarks and gluons is highly suppressed,
a Higgs boson is more likely to be produced through \ac{FSR} or as part of the same process that produces the \ac{DM} particles.
This means that this topology is only sensitive to models where the Higgs boson couples directly to \ac{DM} or some other \ac{BSM} particle involved in \ac{DM} production.
However, in these cases the \ac{DM}--\ac{SM} interaction is probed directly~\cite{Carpenter:2013xra}, potentially providing more information about the structure of the \ac{DM}--\ac{SM} coupling in the event of a discovery.
There are also many models where the coupling between \ac{BSM} particles and the Higgs boson is enhanced, for instance models where \ac{DM} is connected to electroweak symmetry breaking \cite{Bertone:2004pz,Burgess:2000yq} or where \ac{DM} particles couple to the \ac{SM} only through the Higgs sector (Higgs portal models)~\cite{MarchRussell:2008yu}.
These features make the mono-Higgs signature an important part of the \ac{LHC} \ac{DM} search programme.
 
The \ac{2HDM}~\cite{Branco:2011iw} extends the \ac{SM} with a second Higgs doublet.
This predicts a total of five Higgs bosons after mixing: two charged scalars $H^\pm$, two neutral CP-even scalars $h$ and $H$, and one neutral CP-odd scalar $A$.
Two simplified benchmark signal models are used: the $Z'$-2HDM~\cite{Berlin:2014cfa} and 2HDM+$a$~\cite{Abe:2018bpo,Bauer_2017}.
In all models considered here the mass of the lighter neutral CP-even scalar $h$ is required to match that of the Higgs boson observed at the LHC and the Yukawa couplings are defined according to the Type II 2HDM.
 
\begin{figure}[!ht]
\centering
\includegraphics[width=0.4\textwidth]{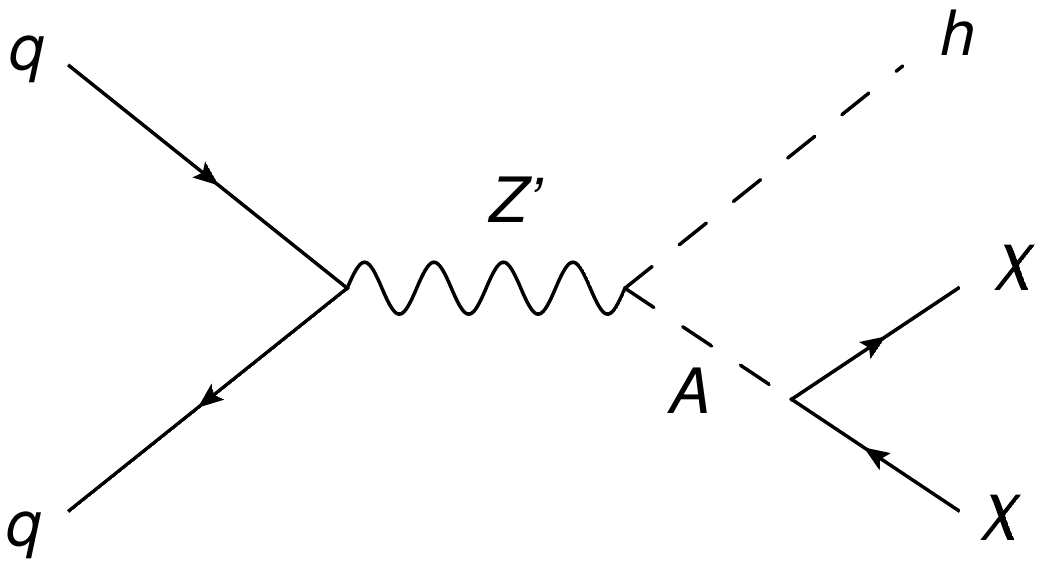}
\caption{Feynman diagram for the production of the mono-Higgs signature in the $Z'$-2HDM.}
\label{fig:FeynZp2HDM}
\end{figure}

The $Z'$-2HDM has an additional heavy vector boson, denoted $Z'$, whose coupling to quarks $g_{Z'}$ and mass $m_{Z'}$ are free parameters.
The production mechanism for the mono-Higgs signature is shown in Figure~\ref{fig:FeynZp2HDM}.
\ac{DM} is introduced into the model as a new fermion which couples to the CP-odd scalar $A$ with a coupling strength denoted by $g_\chi$. The coupling strength and the DM particle mass $m_{\chi}$ are treated as free parameters.
This model is used mainly as a benchmark for high-mass resonances.
 
\begin{figure}[!ht]
\centering
\begin{subfigure}[b]{0.4\textwidth}
\includegraphics[width=\textwidth]{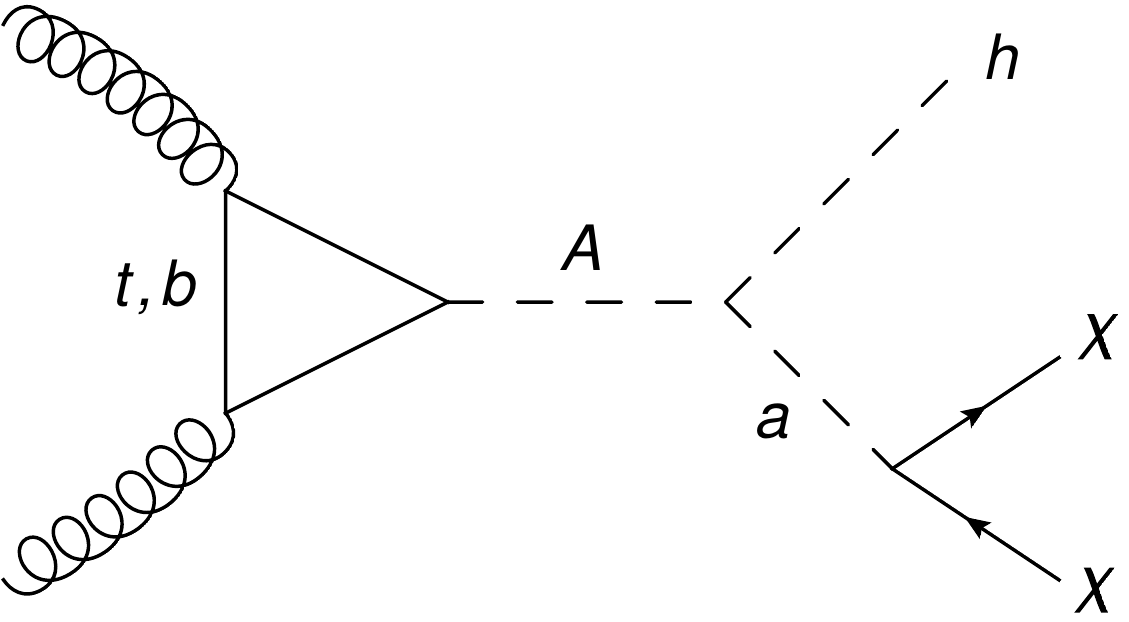}
\caption{}
\label{fig:Feyn2HDMa.ggF}
\end{subfigure}
\begin{subfigure}[b]{0.4\textwidth}
\includegraphics[width=\textwidth]{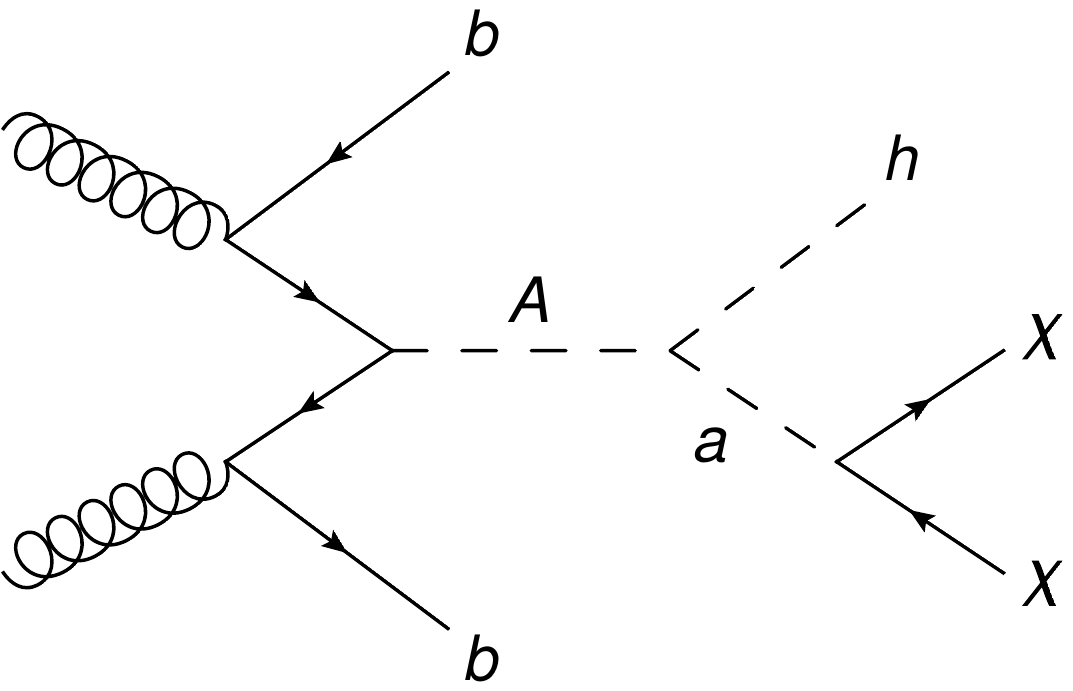}
\caption{}
\label{fig:Feyn2HDMa.bbA}
\end{subfigure}
\caption{Feynman diagrams for the main production mechanisms of the mono-Higgs signature in the 2HDM+$a$ scenario: \subref{fig:Feyn2HDMa.ggF}~gluon--gluon fusion and \subref{fig:Feyn2HDMa.bbA}~$b$-associated production.}
\label{fig:Feyn2HDMa}
\end{figure}

The 2HDM+$a$ scenario is the simplest renormalisable and gauge-invariant extension of a simplified pseudoscalar mediator model.
It adds a new pseudoscalar singlet which mediates the interactions between the \ac{SM} and a singlet fermion $\chi$ identified as the \ac{DM} candidate.
The coupling between the pseudoscalar and \ac{DM} singlets, denoted $y_\chi$, and the mass of the \ac{DM} $m_\chi$ are free parameters.
This singlet mixes with the pseudoscalar $A$ from the two Higgs doublets, with the mixing angle $\theta$ and the mass of the resulting pseudoscalar $a$ being free parameters of the model.
A major advantage of the 2HDM+$a$ scenario over simpler models is that it generates a wider variety of experimental signatures which can provide complementary exclusion regions from different types of experiment.
There are two main production mechanisms for the mono-Higgs signature in this model, as shown in Figure~\ref{fig:Feyn2HDMa}.
 
In the type-II \acs{2HDM} considered in this paper the coupling between down-type quarks and the $A$ boson scales with \tanb, the ratio of the vacuum expectation values of the two Higgs doublets.
This means that for low \tanb\ values ($\tanb \lesssim 5$) the gluon--gluon fusion ($gg$F) mechanism shown in Figure~\ref{fig:Feyn2HDMa.ggF} dominates, whereas for higher \tanb{} values the $b$-associated production ($bbA$) shown in Figure~\ref{fig:Feyn2HDMa.bbA} is dominant.
Signal grids are generated where each of the two production mechanisms are used exclusively.
For each grid a \tanb\ value is chosen that ensures that the corresponding production mechanism is dominant: $\tanb = 1$ for the $gg$F grid and $\tanb = 10$ for the $bbA$ grid.
 
Similar analyses were performed using data taken during the years 2015--2016 by ATLAS~\cite{EXOT-2016-25} and CMS~\cite{CMS-EXO-18-011}.
Other mono-Higgs analyses were also performed on the same datasets in final states where the Higgs boson decays into a pair of photons in ATLAS~\cite{Aad:2021qks} or either a pair of photons, a $\tau^+\tau^-$ pair in CMS~\cite{CMS-EXO-16-055} or a pair of $W$ or $Z$ bosons \cite{CMS-EXO-18-011}.
Beyond the large increase in integrated luminosity (from 36~\ifb\ to 139~\ifb) a number of analysis improvements extend the sensitivity beyond the previous ATLAS search.
 
In the previous ATLAS search, events were required to have either one or two $b$-jets, whereas in this search the events are divided into regions with either exactly two or at least three $b$-jets.
Introducing the exclusive three $b$-jet category improves the sensitivity to the 2HDM+$a$ $bbA$ production mechanism which was not considered in the previous ATLAS search,
while the one $b$-jet category does not provide any significant improvement due to large backgrounds with high uncertainties.
 
The analysis also benefits from using particle-flow objects for jet reconstruction~\cite{PERF-2015-09}, neural-network based $b$-jet~\cite{ATL-PHYS-PUB-2017-013} and $\tau$-lepton~\cite{ATL-PHYS-PUB-2019-033} identification, and variable-radius track-jets~\cite{Krohn:2009zg} to identify boosted Higgs boson candidates.
In order to reduce backgrounds containing fake \MET the likelihood-based \MET significance $\mathcal{S}$~\cite{ATLAS-CONF-2018-038} is used.
The event selections were reoptimised resulting in an improved sensitivity, especially to highly boosted signals.
 
\section{ATLAS detector}
\label{sec:detector}
 
\newcommand{\AtlasCoordFootnote}{
ATLAS uses a right-handed coordinate system with its origin at the nominal interaction point (IP)
in the centre of the detector and the \(z\)-axis along the beam pipe.
The \(x\)-axis points from the IP to the centre of the LHC ring,
and the \(y\)-axis points upwards.
Cylindrical coordinates \((r,\phi)\) are used in the transverse plane,
\(\phi\) being the azimuthal angle around the \(z\)-axis.
The pseudorapidity is defined in terms of the polar angle \(\theta\) as \(\eta = -\ln \tan(\theta/2)\).
Angular distance is measured in units of \(\Delta R \equiv \sqrt{(\Delta\eta)^{2} + (\Delta\phi)^{2}}\).}
 
The ATLAS detector~\cite{PERF-2007-01} at the LHC covers nearly the entire solid angle around the collision point.\footnote{\AtlasCoordFootnote}
It consists of an inner tracking detector surrounded by a thin superconducting solenoid, electromagnetic and hadronic calorimeters,
and a muon spectrometer incorporating three large superconducting toroidal magnets.
 
The inner-detector system (ID) is immersed in a \SI{2}{\tesla} axial magnetic field
and provides charged-particle tracking in the range \(|\eta| < 2.5\).
The high-granularity silicon pixel detector covers the vertex region and typically provides four measurements per track,
the first hit normally being in the insertable B-layer installed before Run~2~\cite{ATLAS-TDR-2010-19,PIX-2018-001}.
It is followed by the silicon microstrip tracker, which usually provides eight measurements per track.
These silicon detectors are complemented by the transition radiation tracker (TRT),
which enables radially extended track reconstruction up to \(|\eta| = 2.0\).
The TRT also provides electron identification information
based on the fraction of hits (typically 30 in total) above a higher energy-deposit threshold corresponding to transition radiation.
 
The calorimeter system covers the pseudorapidity range \(|\eta| < 4.9\).
Within the region \(|\eta|< 3.2\), electromagnetic calorimetry is provided by barrel and
endcap high-granularity lead/liquid-argon (LAr) calorimeters,
with an additional thin LAr presampler covering \(|\eta| < 1.8\)
to correct for energy loss in material upstream of the calorimeters.
Hadronic calorimetry is provided by the steel/scintillator-tile calorimeter,
segmented into three barrel structures within \(|\eta| < 1.7\), and two copper/LAr hadronic endcap calorimeters.
The solid angle coverage is completed with forward copper/LAr and tungsten/LAr calorimeter modules
optimised for electromagnetic and hadronic measurements, respectively.
 
The muon spectrometer (MS) comprises separate trigger and
high-precision tracking chambers measuring the deflection of muons in a magnetic field generated by the superconducting air-core toroids.
The field integral of the toroids ranges between \num{2.0} and \SI{6.0}{\tesla\metre}
across most of the detector.
A set of precision chambers covers the region \(|\eta| < 2.7\) with three layers of monitored drift tubes,
complemented by cathode-strip chambers in the forward region, where the background is highest.
The muon trigger system covers the range \(|\eta| < 2.4\) with resistive-plate chambers in the barrel, and thin-gap chambers in the endcap regions.
 
Interesting events are recorded by the first-level trigger system implemented in custom hardware,
followed by selections made by algorithms implemented in software in the high-level trigger~\cite{TRIG-2016-01}.
The first-level trigger reduces the output event rate from the \SI{40}{\MHz} bunch crossing rate to below \SI{100}{\kHz},
which the high-level trigger further reduces in order to record events to disk at about \SI{1}{\kHz}.
 
An extensive software suite~\cite{ATL-SOFT-PUB-2021-001} is used for real and simulated data reconstruction
and analysis, for operation and in the trigger and data acquisition systems of the experiment.
\section{Data and simulated event samples}
\label{sec:samples}
 
This search uses 139~\ifb{} of proton--proton collision data recorded by the ATLAS detector at a centre-of-mass energy of 13~\TeV{} during the years 2015--2018 (Run~2).
The uncertainty in the total integrated luminosity for the full Run~2 dataset is 1.7\%~\cite{ATLAS-CONF-2019-021}, obtained using the LUCID-2 detector \cite{Avoni_2018} for the primary luminosity measurements.
All events used are required to pass basic data-quality requirements which ensure that all components of the ATLAS detector were functioning correctly~\cite{ATLAS-DAPR-2018-01}.
Events selected for the analysis search regions were collected by the primary \MET triggers~\cite{TRIG-2019-01}, which select those that have a large transverse momentum imbalance.
The algorithms used to calculate the imbalance, and the thresholds required, varied during the data-taking.
The `primary' \MET trigger in a run is the most efficient available \MET trigger, in all cases reaching full efficiency by an offline \MET value of approximately 200~\GeV.
 
Simulated event samples corresponding to the $Z'$-2HDM signal were generated at leading order (LO) in QCD in the 5-flavour scheme using \mgamc~v2.6.5~\cite{Alwall:2014hca} interfaced to \pythia.240~\cite{Sjostrand:2014zea} using a set of tuned parameters called the A14 tune~\cite{ATL-PHYS-PUB-2014-021}.
Following the recommendations of the ATLAS-CMS Dark Matter Forum~\cite{Abercrombie:2015wmb} the coupling of the $Z'$ boson to quarks was fixed to $g_{Z'} = 0.8$, the mass of the DM candidate was set to $m_{\chi} = 100$~\GeV, the $A\chi\bar{\chi}$ coupling $g_\chi$ was set to 1, $\tan\beta$ was set to 1, and the alignment limit, i.e.\ $\sin(\beta-\alpha) = 1$, was assumed, where $\alpha$ is the mixing angle between the two CP-even Higgs bosons.
The samples were generated with varying $m_{Z'}$ between 600 and 3600\,\GeV\ and $m_A$ between 300 and 1300\,\GeV.
 
Simulated event samples corresponding to the 2HDM+$a$ signal were generated at LO using \mgamc~v2.6.7 interfaced to \pythia.244 with the A14 tune.
Samples were generated separately for the $gg$F and $bbA$ production modes, with the former generated in the 4-flavour scheme setting $\tanb = 1$ and the latter in the 5-flavour scheme setting $\tanb = 10$.
Following the recommendations of the LHC Dark Matter Working Group~\cite{Abe:2018bpo}, the mass of the DM candidate was set to $m_{\chi}=10$~\GeV, the Yukawa coupling between the DM candidate and the pseudoscalar $a$ was set to $y_{\chi}=1$ and the Higgs quartic couplings were set to $\lambda_3=\lambda_{P1}=\lambda_{P2}=3$.
The chosen value of $m_\chi$ ensures that the $a\rightarrow\chi\overline{\chi}$ branching ratio is significant for all values of $m_a$ used.
The pseudoscalar mixing angle was set to $\sin\theta=0.35$ and the alignment limit was assumed.
The samples were generated with varying $m_A$ between 250 and 2000\,\GeV\ and $m_a$ between 100 and 600\,\GeV.
 
For all signal samples the masses of the heavy Higgs bosons were considered degenerate ($m_A = m_H = m_{H^\pm}$).
The mass of the lightest CP-even Higgs boson was set to match that of the Higgs boson discovered at the LHC, i.e.\ $m_h = 125\,\GeV$~\cite{HIGG-2014-14}.
 
Background events from the production of a single weak vector boson ($V=W,Z$) in association with jets or of a pair of weak bosons (diboson) were simulated using \sherpa~2.2.1~\cite{Bothmann:2019yzt}, with \sherpa~2.2.2 used for $gg$-initiated diboson production.
For the $V$+jets samples, \ac{NLO} matrix elements for up to two jets and LO matrix elements for up to four jets were calculated using the \comix~\cite{Gleisberg:2008fv} and \textsc{OpenLoops}~\cite{Cascioli:2011va,Denner:2016kdg} libraries.
For the $q\bar{q}$-initiated diboson samples, \ac{NLO} matrix elements for up to one additional jet and LO matrix elements for up to three additional jets were used, while the $gg$-initiated processes were generated using LO matrix elements for up to one additional jet.
The samples were matched with the \sherpa parton shower~\cite{Schumann:2007mg} using the \mepsatnlo prescription~\cite{Hoeche:2011fd,Hoeche:2012yf,Catani:2001cc,Hoeche:2009rj}.
 
Samples corresponding to the top-quark pair (\ttbar), single-top-quark (s-, t- and $Wt$-channels) and \ttH processes were generated using \powhegbox~v2~\cite{Frixione:2007nw,Nason:2004rx,Frixione:2007vw,Alioli:2010xd,Re:2010bp,Frederix:2012dh,Alioli:2009je,Hartanto:2015uka} with $h_\mathrm{damp}$ set to $1.5\mtop$ and $\mtop=172.5\,\GeV$.
The $h_\mathrm{damp}$ parameter regulates the transverse momentum (\pT) of the high-\pT emission against which the \ttbar system recoils.
The inclusive cross-section for these processes were corrected to \ac{NNLO} plus \ac{NNLL} accuracy for \ttbar~\cite{Beneke:2011mq,Cacciari:2011hy,Baernreuther:2012ws,Czakon:2012zr,Czakon:2012pz,Czakon:2013goa,Czakon:2011xx}, to \ac{NLO}+NNLL accuracy for $Wt$~\cite{Aliev:2010zk,Kant:2014oha}, to \ac{NLO} accuracy for s- and t-channel single top-quark production~\cite{Aliev:2010zk,Kant:2014oha} and to \ac{NLO} QCD+\ac{EW} accuracy for $\ttH$~\cite{deFlorian:2016spz}.
The diagram removal scheme~\cite{Frixione:2008yi} was used in the $Wt$ samples to avoid double counting contributions from \ttbar processes.
 
The $W$/$Z$+$h$ samples were generated using \powhegbox~v2.
The cross-sections of the $q\bar{q}$-initiated processes were calculated at \ac{NNLO} QCD and \ac{NLO} \ac{EW} accuracy, using the \minlo procedure~\cite{Hamilton:2012np,Luisoni:2013kna}.
The cross-sections of the $gg$-initiated processes were calculated at \ac{NLO}+\ac{NLL} accuracy in QCD~\cite{Altenkamp:2012sx,Harlander:2014wda,Brein:2012ne}.
 
The $\ttV$ samples were generated using \mgamc~v2.3.3 at \ac{NLO}.
The cross-sections were calculated at \ac{NLO} QCD and \ac{EW} accuracies as provided by Ref.~\cite{deFlorian:2016spz}.
 
All samples were generated using the \nnpdfnlo \ac{PDF} set~\cite{Ball:2014uwa} apart from the \sherpa $W$/$Z$+jets and diboson samples, which were generated using the \nnpdfnnlo \ac{PDF} set along with a dedicated tune developed by the \sherpa authors, and the t-channel single-top-quark production samples, which were generated using the \nnpdfnlof \ac{PDF} set.
The \powhegbox samples were interfaced with \pythia.230 for the parton shower and hadronisation.
Of these, the \ttbar, single-top-quark and \ttH samples used the \nnpdftwo \ac{PDF} set~\cite{Ball:2014uwa} and the A14 tune~\cite{ATL-PHYS-PUB-2016-020}, while the $W$/$Z$+$h$ samples used the \cteq \ac{PDF} set~\cite{Pumplin:2002vw} and the AZNLO tune~\cite{STDM-2012-23}.
For the \powhegbox top-quark samples the decays of $b$- and $c$-hadrons were simulated using \evtgen~1.6.0~\cite{Lange:2001uf}.
The \ttV samples were interfaced with \pythia.210 using the same tune and \ac{PDF} set as the \powhegbox \ttbar, single-top-quark and \ttH samples, and using \evtgen~1.2.0 for the decays of $b$- and $c$-hadrons.
 
In order to simulate the effect of additional $pp$ collisions in the same and neighbouring bunch crossings (pile-up) all samples were overlaid with multiple $pp$ collisions simulated with \pythia.186 using the \nnpdftwo \ac{PDF} set and the A3 tune~\cite{ATL-PHYS-PUB-2016-017}.
The response of the detector was modelled with a detector simulation~\cite{SOFT-2010-01} based on \textsc{Geant4}~\cite{Agostinelli:2002hh}.
\section{Object definitions}
\label{ssec:selections.objects}
\paragraph{Primary Vertex}\mbox{}
 
Primary vertices are constructed using at least two ID tracks with $\pT > 500\,\MeV$~\cite{ATL-PHYS-PUB-2015-026}.
The primary vertex with the largest sum of squared track transverse momenta ($\sum \pT^2$) is selected as the hard-scatter vertex, henceforth only referred to as \emph{the} primary vertex.
 
\paragraph{Jets}\mbox{}
 
Jets are reconstructed using the \antikt\ algorithm \cite{Cacciari:2008gp,Fastjet}.
The analysis considers three types of jets to better match the different event topologies.
Small-radius (small-$R$) jets are constructed from particle-flow objects formed from ID tracks and calorimeter energy clusters~\cite{PERF-2015-09} using a radius parameter of $R=0.4$.
This radius parameter is designed to capture jets initiated by a gluon, light quark or $b$-quark.
Small-$R$ jets are classified as central ($|\eta| < 2.5$) or forward ($2.5 < |\eta| < 4.5$).
Central small-$R$ jets are required to have $\pT > 20\,\GeV$ and forward small-$R$ jets $\pT > 30\,\GeV$.
In order to remove the impact of jets predominantly formed from particles from pile-up vertices, central small-$R$ jets are required to pass the `Tight' jet vertex tagger (JVT)~\cite{PERF-2014-03} \ac{WP}.\footnote{The JVT selection is applied to jets with $20\,\GeV < \pT < 60$\,\GeV\ and $|\eta|<2.4$.}
 
In topologies where the Higgs boson decay $h\rightarrow b\bar{b}$ cannot be resolved into two small-$R$ jets, large-radius (large-$R$) jets with a radius parameter of $R=1.0$ are used, constructed from calorimeter energy clusters calibrated using the \ac{LCW} scheme~\cite{PERF-2014-07}.
This radius parameter is chosen so that a single large-$R$ jet should capture all jets produced in the decay of a boosted heavy object, such as a Higgs boson.
To reduce the impact of pile-up, these jets are then `trimmed', removing any $R=0.2$ subjets which have less than 5\% of the original jet energy~\cite{Krohn:2009th}.
In order to identify subjets originating from $b$-hadrons within the large-$R$ jets, jets are also constructed from ID tracks, using a variant of the \antikt\ algorithm with a radius parameter that shrinks as the \pT of the proto-jet increases~\cite{Krohn:2009zg}.
These are referred to as variable-radius (variable-$R$) track-jets and are matched to the large-$R$ jets by ghost association~\cite{Cacciari:2008gn}.
The radius parameter is set to $R=30\,\GeV/\pT$, with minimum and maximum values of 0.02 and 0.4, respectively.
The reduced radius at high \pT allows the algorithm to reconstruct separate jets from closely spaced $b$-hadrons, such as in highly boosted $h\rightarrow b\bar{b}$ decays.
 
Both the small-$R$ and large-$R$ jet energies are calibrated using a sequence of simulation-derived corrections.
Small-$R$ jets additionally have an area-based energy subtraction applied to reduce the impact of pile-up, as well as a series of additional data-derived corrections~\cite{Aad:2020flx}.
 
Central small-$R$ jets and variable-$R$ track-jets containing $b$-hadrons are identified using the DL1 tagger~\cite{ATL-PHYS-PUB-2017-013}.
This multivariate algorithm uses the impact parameters of ID tracks as well as information about secondary vertices and reconstructed flight paths of $b$- and $c$-hadrons within the jet.
For both classes of jet a \ac{WP} is chosen which tags jets containing $b$-hadrons with 77\% efficiency in \ttbar events.
The decays of these $b$-hadrons can produce muons which are vetoed when building particle-flow objects and therefore not included in the energies of either the small- or large-$R$ jets.
In order to correct for this, the four-momenta of non-isolated muons falling inside these jet cones can be added into the jet, improving the resolution of their four-momenta.
For small-$R$ (large-$R$) jets, this is done for the muon (two muons) closest to the jet axis.
This correction is only used when calculating the mass of the Higgs boson candidate $m_h$ and was shown in Ref.~\cite{HIGG-2016-29} to improve the resolution of this measurement.
Correcting $m_h$ in this way improves the ability of the fit to separate the signal from the major backgrounds.
 
\paragraph{Leptons}\mbox{}
 
Leptons are divided into `baseline' and `signal' categories.
Events with baseline leptons are vetoed in the analysis search regions and signal leptons are used to define control regions to constrain background components.
 
Electrons are reconstructed from a track which is coincident with a cluster built from energy deposits in the calorimeter~\cite{EGAM-2018-01}.
They are then identified using a multivariate likelihood technique, using several features including the shape of the measured shower, the track quality and the distribution of energy within the calorimeter~\cite{PERF-2017-01}.
For this analysis, the `LooseAndBLayer' \ac{WP}~\cite{PERF-2017-01} is used for both the signal and baseline electrons.
Isolation selections are also applied to distinguish between electrons produced in the initial collision or decays of $W/Z$ bosons or $\tau$-leptons (prompt) and those produced in decays of other objects~\cite{PERF-2017-01}.
Requirements are placed on the energy of calorimeter clusters and the \pT of tracks measured in isolation cones around the electron.
For signal electrons with $\pT < 200~\GeV$ and all baseline electrons the total energy of clusters within $\Delta R = 0.2$ of the electron, excluding the electron cluster, must be less than 20\% of the \pT\ of the electron.
The total \pT\ of tracks matched to the primary vertex that lie within a cone whose size is set to the smaller of $\Delta R=10\,\GeV/\pT$ and 0.2, excluding the electron track, must be less than 15\% of the \pT\ of the electron.
For signal electrons with $\pT > 200~\GeV$ the total energy of clusters within $\Delta R =0.2$ of the electron is required to be less than the smaller of  $0.015 \times \pT$ and $3.5\,\GeV$.
For these electrons, no track-based isolation selection is applied.
Both the baseline and signal electrons are required to have $|\eta| < 2.47$.
Baseline electrons are required to have $\pT > 7~\GeV$ and signal electrons $\pT > 27~\GeV$.
In order to ensure that they are compatible with the primary vertex, the track from which the electron is reconstructed is required to have $\sigma(d_0) < 5$ and
$|z_0\sin\theta| < 0.5$~mm, where $\sigma(d_0)$ is the significance of the transverse impact parameter, $z_0$ is the longitudinal impact parameter, and $\theta$ is the polar angle of the track.
 
Muons are reconstructed by matching track segments formed in the MS to a track from the ID~\cite{MUON-2018-03}.
Identification is performed through selections on the qualities of the tracks used in the reconstruction, as well as their compatibility, for example, in the measurements of \pT in the MS and ID.
Similarly to electrons, isolation selections are also applied.
Baseline muons are required to pass the `Loose' identification \ac{WP} and signal muons are required to pass the `Medium' identification \ac{WP}~\cite{MUON-2018-03}.
For baseline muons, the total energy of clusters within $\Delta R = 0.2$ of the muon is required to be less than 30\% of the \pT\ of the muon.
For signal muons, the total \pT\ of tracks within $\Delta R = 0.2$ of the muon's primary track is required to be less than 1.25\,\GeV.
Both the baseline and signal muons are required to satisfy $|\eta| < 2.5$, $\sigma(d_0) < 3$ and $|z_0\sin\theta| < 0.5$~mm.
Baseline muons are required to have $\pT > 7~\GeV$ and signal muons $\pT > 25~\GeV$.
 
Hadronically decaying $\tau$-lepton reconstruction is seeded from $R=0.4$ \antikt jets built using the \ac{LCW}-calibrated clusters~\cite{ATL-PHYS-PUB-2015-045}.
As hadronic $\tau$-lepton decays yield either one or three charged pions the jets are required to have either one or three tracks within $\Delta R = 0.2$ of the jet axis.
A \ac{RNN} classifier is used to identify the $\tau$-leptons~\cite{ATL-PHYS-PUB-2019-033}.
The inputs to the \ac{RNN} are built from the clusters and tracks associated with the $\tau$-lepton.
All $\tau$-leptons are required to pass the `VeryLoose' \ac{WP}~\cite{ATL-PHYS-PUB-2019-033} and have $|\eta| < 2.5$ and $\pT > 20~\GeV$.
As there is no dedicated $\tau$-lepton control region, no signal $\tau$-lepton selection is defined.
 
\paragraph{Overlap removal}\mbox{}
 
In order to avoid the same detector signals being interpreted as different objects, an overlap removal procedure is applied as follows.
If any object is rejected at one step it is not considered in later steps.
First, if any two electrons share a track the electron with the lower \pT\ is removed.
Next, any $\tau$-leptons within $\Delta R = 0.2$ of an electron or muon are removed.
Then, any electrons which share a track with a muon are removed.
If any small-$R$ jet is within $\Delta R = 0.2$ of an electron it is removed, and then any electron within a cone of \pT-dependent size around a small-$R$ jet is removed.
If any small-$R$ jet with fewer than three tracks has an associated muon or is within $\Delta R = 0.2$ of one it is removed, and then any muons within a cone of \pT-dependent size around a small-$R$ jet are removed.
Next, any small-$R$ jets within $\Delta R = 0.2$ of a $\tau$-lepton are removed.
Finally, any large-$R$ jets within $\Delta R = 1.0$ of an electron are removed.
 
Track-jets do not participate in the overlap removal as they are only used for $b$-tagging.
 
\paragraph{Missing transverse momentum}\mbox{}
 
The missing transverse momentum (with magnitude \MET~\cite{ATLAS-CONF-2018-023}) is defined as the negative vector sum of the transverse momenta of all the observable objects in the event, plus a soft term including ID tracks matched to the primary vertex but not to any of the other objects.
The \MET reconstruction uses the baseline electrons and muons as well as all small-$R$ jets, and employs
a separate overlap removal procedure which takes into account detector signals from each object included~\cite{ATLAS-CONF-2018-023}.
In control regions a modified definition of \MET\ is used in which electrons and muons are treated as invisible, \METnolep, to imitate the kinematics of the $Z\to\nu\bar{\nu}$ background process.
 
Object mismeasurements, especially of jets, are the main source of fake \MET.
Therefore, the \MET significance (\METSig)~\cite{ATLAS-CONF-2018-038} is defined to assess the likelihood that the \MET\ is really due to invisible particles or is more likely to come from mismeasurements.
It is calculated using the expected resolutions of all objects which enter the \MET calculation and the correlations between them.
 
\section{Event selection}
\label{sec:selections}

The basic target final-state topology is a Higgs boson decaying into two $b$-quarks produced with a significant imbalance in the measured transverse momentum.
Events are divided into non-overlapping regions designed either to be enriched in the signal process (signal regions) or in a significant background process (control regions).
Control regions differ from signal regions primarily through requiring the presence of one or two lepton(s), whereas signal regions veto events containing baseline leptons.
 
As the angle between the two $b$-jets produced in the Higgs boson decay is inversely proportional to the \pT of the Higgs boson, in cases where the Higgs boson is significantly boosted it can become difficult to reconstruct the two $b$-quarks as separate jets.
This motivates splitting the analysis into `resolved' regions in which the decay products of the Higgs boson are reconstructed as two separate jets and `merged' regions in which the entire Higgs boson decay is reconstructed as a single jet.
 
In $b$-associated production within the 2HDM+$a$ benchmark model, the Higgs boson and \ac{DM} particles are produced with an extra pair of $b$-quarks from gluon splitting.
Therefore, to enhance sensitivity to these models, all regions are further split into those requiring exactly two $b$-jets and those requiring $\geq$3 $b$-jets (referred to as \2b and \3b, respectively).
 
\subsection{Common selections}
 
Events which do not have a reconstructed primary vertex are rejected.
Events are also rejected if found to contain any jets with properties consistent with beam-induced backgrounds, cosmic-ray showers or noisy calorimeter cells~\cite{ATLAS-CONF-2015-029}.
 
Events are required to have $\MET > 150~\GeV$ and are vetoed if they contain a baseline $\tau$-lepton.
In order to further reduce the background from $\tau$-lepton decays, events are also vetoed if they have any small-$R$ jets with $\Delta\phi(\mathrm{jet},\,\MET) < 22.5\degree$ where the track multiplicity in the jet is between 1 and 4.
This selection is referred to as the `extended $\tau$-lepton veto'.
A further source of background is \MET arising from either leptonic heavy-flavour decays in a jet or a jet which is severely mismeasured.
In these cases, the \MET tends to be aligned with the jet; therefore, events where any of the up to three leading small-$R$ jets have $\Delta\phi(\mathrm{jet},\,\MET) < 20\degree$ are rejected.
For control regions, these requirements use $\METnolep$ rather than $\MET$.
 
Only loose selections are placed on the mass of the Higgs boson candidate ($m_h$, defined in the following sections) because it is used as the discriminating variable for the final fit.
The range is $50\,\GeV < m_h < 280\,\GeV$ for the resolved regions and $50\,\GeV < m_h < 270\,\GeV$ for the merged regions, with the lower limit chosen to be the lowest calibrated large-$R$ jet mass and the upper limit chosen to be significantly larger than the Higgs boson mass, with the precise value being determined by the $m_h$ binning used in the fit, which depends on the available sample size.
 
\subsection{Signal regions}
\label{ssec:selections.SRs}
 
The signal region selections for both the merged and resolved regions are summarised in Table~\ref{tab:selections.summary}.
All signal region events are required to have passed the primary \MET trigger~\cite{TRIG-2019-01}.
In order to reduce the contribution of \ac{SM} processes producing \MET through the decay $W\rightarrow\ell\nu$, events are rejected if they contain any baseline electron or muon.
 
\subsubsection{Resolved regions}
\label{sssec:selections.resolved}
 
The resolved regions are defined by selecting events with $\MET < 500\,\GeV$.
Events in the resolved regions are required to have at least two $b$-tagged small-$R$ jets, with the two with the highest \pT forming the Higgs boson candidate.
The combined \pT of this two-jet system (${\pT}_h$) is required to be greater than 100\,\GeV\ and its mass is corrected for nearby muons as described in Section~\ref{ssec:selections.objects} to form $m_h$.
 
The dominant background in the resolved region is \ttbar production where one top quark decays leptonically, but the lepton is either not reconstructed or not correctly identified.
In these cases, all the \MET in the event (beyond that from mismeasurement) originates from the decay of one of the two $W$ bosons, and therefore the transverse mass of the \MET and the corresponding $b$-jet should be approximately bounded from above by the top-quark mass. Here the transverse mass is defined as
\begin{equation*}
m_\mathrm{T}^{b,\mathrm{min}/\mathrm{max}} = \sqrt{2\pT^{b,\mathrm{min}/\mathrm{max}}\MET(1-\mathrm{cos}\Delta\phi(\pT^{b,\mathrm{min}/\mathrm{max}}, \MET))}
\end{equation*}
where $\pT^{b,\mathrm{min}}$ and $\pT^{b,\mathrm{max}}$ are respectively defined as the \pT of the $b$-jet which is closest to (min) or furthest from (max) the \MET\ in $\phi$.
Events are required to satisfy $m_\mathrm{T}^{b,\mathrm{min}} > 170~\GeV$ and $m_\mathrm{T}^{b,\mathrm{max}} > 200~\GeV$.
 
In order to suppress contributions from multijet backgrounds, the object-based \MET\ significance is required to satisfy $\METSig > 12$.
After this selection, data-driven estimates of the remaining multijet contribution to the signal regions were found to be substantially smaller than the expected statistical uncertainty of the data so the impact of multijet processes is not included in the background estimation.
The studied signal models typically have fewer reconstructed jets than the dominant backgrounds.
Therefore, events with exactly two $b$-tagged jets (\2b) are required to have at most four small-$R$ jets, where only central jets are counted.
For events with at least three $b$-tagged jets (\3b), this requirement is relaxed to at most five small-$R$ jets to ensure a sufficient sample size in the corresponding control regions.
 
The \2b\ and \3b\ selections are split into three \MET bins: $150\,\GeV < \MET < 200\,\GeV$, $200\,\GeV < \MET < 350\,\GeV$ and $350\,\GeV < \MET < 500\,\GeV$.
In the highest \MET bin the requirement on ${\pT}_h$ is tightened to being greater than 300 GeV.
This leads to six resolved signal regions: one corresponding to each combination of \MET bin and number of $b$-tagged jets.
 
The \MET triggers become fully efficient at an offline \MET value close to 200~\GeV, but the analysis also uses events in the range $150\,\GeV < \MET < 200~\GeV$.
In order to correct for \ac{MC} mismodelling of the \MET trigger response, the trigger efficiency must be measured in both data and simulation and scale factors calculated to correct the simulation.
Given that the \MET triggers use calorimeter information only, muons are treated as almost invisible particles, meaning that \MET trigger efficiencies can be measured using events selected by single-muon triggers~\cite{TRIG-2018-01}.
The scale factors are calculated as a function of $E_{\mathrm{T,~lep,~invis}}^{\mathrm{miss}}$ in a region whose selection matches the $150\,\GeV < \MET < 200~\GeV$ and \2b\ region except that all \MET selections are dropped, exactly one $b$-tagged jet is required and exactly one signal muon is required.
Events containing electrons are still vetoed.
The scale factors have values in the range 0.95--1.0.

\subsubsection{Merged regions}
\label{sssec:selections.merged}
 
The merged regions are defined by selecting events with $\MET > 500~\GeV$.
At least one large-$R$ jet is required, and the two leading variable-$R$ track-jets associated with the leading large-$R$ jet are required to be $b$-tagged.
This large-$R$ jet is defined to be the Higgs boson candidate and its mass is corrected for nearby muons as described in Section~\ref{ssec:selections.objects} to form $m_h$.
Events are separated into those which have no additional $b$-tagged variable-$R$ track-jets (\2b\ selection)
and those which have at least one such $b$-tagged variable-$R$ track-jet not associated with the Higgs boson candidate (\3b\ selection).
 
The merged \2b\ selection is split into two \MET bins,  $500\,\GeV < \MET < 750~\GeV$ and $\MET > 750~\GeV$, while no further splitting is done for the \3b\ selection.
 
\begin{table}
\centering
\caption{
Summary of selections used to define the signal regions used in the analysis.
The kinematic variables are defined in the text.
}
\vspace{0.2cm}
\begin{tabular}{c|c}
\hline
\textbf{Resolved} & \textbf{Merged} \\
\hline
\multicolumn{2}{c}{Primary \MET trigger} \\
\hline
\multicolumn{2}{c}{Data quality selections} \\
\hline
\multicolumn{2}{c}{\MET > 150 \GeV} \\
\hline
\multicolumn{2}{c}{Lepton veto \& extended $\tau$-lepton veto} \\
\hline
\multicolumn{2}{c}{$\Delta\phi(\mathrm{jet}_{1,2,3}, \MET) > 20\degree$} \\
\hline
\MET < 500 \GeV & \MET > 500 \GeV \\
\hline
At least 2 small-$R$ jets & At least 1 large-$R$ jet \\
\hline
At least 2 $b$-tagged small-$R$ jets & At least 2 $b$-tagged associated variable-$R$ track-jets \\
\hline
${\pT}_h > 100\,\GeV$ if \MET < 350 \GeV & \multirow{2}{*}{---} \\
${\pT}_h > 300\,\GeV$ if \MET > 350 \GeV & \\
\hline
$m_\mathrm{T}^{b,\mathrm{min}} > 170\,\GeV$ & --- \\
\hline
$m_\mathrm{T}^{b,\mathrm{max}} > 200\,\GeV$ & --- \\
\hline
\METSig > 12 & --- \\
\hline
$N_{\textrm{small-}R\,\textrm{jets}} \leq 4$ if \2b & \multirow{2}{*}{---} \\
$N_{\textrm{small-}R\,\textrm{jets}} \leq 5$ if \3b &\\
\hline
$50\,\GeV < m_h < 280\,\GeV$ & $50\,\GeV < m_h < 270~\GeV$ \\
\hline
\end{tabular}
\label{tab:selections.summary}
\end{table}
\subsection{Background modelling and control regions}
\label{sec:backgrounds}
 
The dominant backgrounds in the signal regions consist of \ttbar and $W/Z$ bosons produced in association with heavy-flavour jets. The $W/Z$+jets backgrounds are subdivided according to the true flavour of the jets that constitute the Higgs boson candidate.\footnote{Simulated jets are labelled according to which hadrons with $\pT > 5$~\GeV\ are found within a cone of size $\Delta R = 0.3$ around the jet axis. If a $b$-hadron is found the jet is labelled as a $b$-jet. If no $b$-hadron is found, but a $c$-hadron is present, then the jet is labelled as a $c$-jet. Otherwise the jet is labelled as a light jet. The flavour of the two leading $b$-tagged track-jets is used in the merged region.}  If the flavour of either (or both) those two jets is a $b$-quark, the
event is considered to be $W/Z$+HF background, where HF stands for Heavy Flavour. In the \2b\ resolved regions the dominant backgrounds are \ttbar and $Z$+HF, with the latter becoming more important as the \MET\ increases. The \2b\ merged regions are dominated by $Z$+HF.
Both the resolved and merged \3b\ regions are dominated by \ttbar, where the extra $b$-jet typically is a mis-tagged jet originating from a hadronic $W$ boson decay.
At higher \MET values the $Z$+HF background becomes important again.

The \ttbar, $W$+HF and $Z$+HF contributions are modelled using simulation with their normalisations corrected from data by using background-enriched control regions besides the signal regions.
Smaller backgrounds are taken directly from simulation.
These include the production of a $W$ or $Z$ boson in association with light jets or at most one jet containing a $c$-hadron, single top-quark production (dominated by production in association with a $W$ boson) and diboson processes.
Small contributions also arise from \ttbar\ processes in association with vector bosons or a Higgs boson.
Another background contribution stems from vector-boson production in association with a Higgs boson ($Vh$), which mimics the signal due to the presence of a Higgs boson peak in association with jets.
In the case where the vector boson is a $Z$ boson decaying to neutrinos this is an irreducible background.
Similarly, the diboson decay $ZZ \rightarrow b\overline{b} \nu \nu$ is nearly irreducible due to small difference in $Z$ and $h$ mass peaks (compared to the Higgs candidate mass resolution).
The multijet background is negligible in all regions after the requirements on the object-based \MET\ significance $S$ and on $\Delta \phi(\text{jet},\MET)$, and thus not further considered.
The total background estimates in all regions, including uncertainties, are determined in a simultaneous fit to all regions, which is described in Section~\ref{sec:fit}.
 
Top-quark pair production and $W$+HF processes contribute in the signal regions if leptons in the decays are either not identified or outside the kinematic acceptance. The main contribution arises from decays involving hadronically decaying $\tau$-leptons. As the shape of the event variables is the same for all lepton flavours, the kinematic phase space of the signal region can be closely approximated by control regions requiring an isolated signal muon (1-muon control regions). In this case, to better approximate the signal regions which veto the presence of any leptons, \METnolep\ is used as proxy for \MET. Also, any other variable using \MET\ in its calculation, e.g.\ \MET\ significance and $m_\mathrm{T}^{\mathrm{b,min/max}}$, is constructed using \METnolep. This ensures that the \MET-related quantities in the control regions correspond to those in the signal regions. Otherwise, the 1-muon control regions are defined by the same criteria as the signal regions.
 
As the momentum of the $Z$ boson does not depend on its decay mode, the main background in the signal regions, $Z \rightarrow \nu\nu$ in association with heavy-flavour jets, can be closely modelled by $Z \rightarrow \ell^{+}\ell^{-}$ events.
This means that the normalisation of $Z\rightarrow\nu\nu$+HF contribution can be corrected by measuring $Z\rightarrow\ell^+\ell^{-}$+HF events.
To select these events, control regions requiring exactly two baseline electrons or muons with opposite charge are defined (2-lepton control regions).
These events are collected using primary triggers selecting an isolated electron or muon~\cite{TRIG-2018-05,TRIG-2018-01}.
Further, one of the electrons or muons is required to be a signal electron or muon with $\pT > 27~\GeV$ or $\pT > 25~\GeV$, respectively.
The invariant mass of the leptons is required to be consistent with the mass of the $Z$ boson within 10~\GeV.
While keeping all other criteria of the signal regions, an additional criterion of $S < 5$ is imposed to suppress a remaining contribution of \ttbar\ processes.
To be similar to the signal regions, \METnolep\ is used as proxy for \MET\ and in the calculation of any other variable using \MET.
\section{Statistical analysis}
\label{sec:fit}
 
A binned profile likelihood fit~\cite{RooFit,moneta2011roostats} is used to obtain background estimates and check the compatibility of the data with the background-only hypothesis as well as to extract upper limits at 95\% confidence level (CL) on the signal cross-section. The likelihood function is constructed from a product of Poisson probability functions based on the expected signal and background yields in every region considered in the fit, as shown in Table~\ref{tab:fitChannels}. It contains the parameter of interest, $\mu$, which multiplies the signal cross-section, as well as floating normalisation factors controlling the background normalisations. Systematic uncertainties are included in the likelihood by nuisance parameters (NP) $\theta$ which are parameterised by Gaussian or log-normal priors.
 
Four normalisation factors are used for the backgrounds, scaling the \ttbar\ and $W/Z$+HF processes. Of those, two normalisation factors are used for $Z$ boson production in association with two $b$-tagged jets or at least three $b$-tagged jets. In the \2b\ and \3b\ selections, \ttbar\ and $W$+HF are normalised by a single parameter each, as the production mechanism for \ttbar\ is the same in the two categories and the contribution of $W$+HF is minor in the \3b\ selection.
 
The likelihood includes all control and signal regions, as detailed in Table~\ref{tab:fitChannels}, where these regions are binned to increase the sensitivity and improve the determination of the normalisation factors for \ttbar\ and $W/Z$+HF.
The signal regions are binned in the invariant mass of the Higgs boson candidate as defined in Section~\ref{sec:selections}. The chosen binning in the signal regions is optimised to obtain the best expected sensitivity to the signal models addressed in this paper, while also keeping statistical uncertainties in each bin low.
 
The 1-muon control regions are split in positive or negative muon charge in the case of the \2b\ regions. This improves the separation between \ttbar\ and $W$+HF processes, as $W$+HF processes exhibit a charge asymmetry in $pp$ collisions. Only inclusive event yields are used in the other regions, which are dominated by \ttbar, and for the 2-lepton control regions.

\begin{table}[tbh]
\centering
\caption{Event categories entering the combined fit of the model to the data. The discriminant $m_{h}$ denotes the mass of the light Higgs boson candidate and corresponds either to the dijet mass $m_{jj}$ in the regions selecting two small-$R$ jets or to the large-$R$ jet mass $m_{J}$ for the regions requiring a large-$R$ jet. ``Yields" refers to the number of events in a given region.}
\vspace{0.2cm}
\small
\begin{center}
\begin{tabular}{l|c|c|c}
\hline
& \textbf{0 lepton}     & \textbf{1 muon}                        & \textbf{2 leptons}            \\
\hline
Aim                               & Signal regions       & $t\bar t$ and $W$+HF control region  & $Z$+HF control regions         \\\hline
\multirow{2}{*}{Fitted observable}         & \multirow{2}{*}{$m_{h}$ distribution}  & Muon charge (\2b)  & \multirow{2}{*}{Yields}   \\
&                & Yields (\3b) & \\\hline
 
\multirow{2}{*}{$b$-tag multiplicities}            &                           \multicolumn{3}{c}{resolved (small-$R$ jets): 2, $\geq$3}                                       \\
& \multicolumn{3}{c}{merged (variable-$R$ track-jets):}  \\
&                                    
\multicolumn{3}{c}{2 (inside $h$ candidate), $\geq$3 (2 inside $h$ candidate)}  \\\hline
 
\MET proxy & \met                  & \METnolep     & \METnolep            \\\hline
\multirow{3}{*}{Bins in \MET\ proxy}&               \multicolumn{3}{c}{resolved:\ \ [150, 200),\ \ [200, 350)\ \ and\ \ [350, 500)~\GeV}       \\\cline{2-4}
&                          \multicolumn{3}{c}{\2b\ merged signal regions (0 lepton):\ \  [500, 750)\ \ and\ \ [750, $\infty$)~\GeV}                             \\
&                          \multicolumn{3}{c}{Other merged regions:\ \  [500, $\infty$)~\GeV}                             \\
\hline
\end{tabular}
\end{center}
\label{tab:fitChannels}
\end{table}

The nominal fit results
are obtained by maximising the likelihood
function with respect to all parameters.
Two different fit configurations are used: in the background-only profile likelihood fit, the parameters are determined in a fit to data assuming the presence of no signal. The second fit configuration instead allows for the presence of a specific signal, and is referred to as the model-dependent fit.
 
The test statistic $q_\mu$ is constructed using the profile likelihood~\cite{Cowan:2010js}: $q_\mu = - 2
\ln (\mathcal{L} (\mu, \hat{\hat{\theta_\mu}})/\mathcal{L} (\hat{\mu}, \hat{\theta}))$,
where $\hat{\mu}$ and $\hat{\theta}$ are the parameters that maximise the likelihood, and $\hat{\hat{\theta}}_\mu$ are the nuisance parameter
values that maximise the likelihood for a given $\mu$. This test statistic is used to measure the
compatibility of the background-only model with the observed data and to derive exclusion intervals
using the CL$_{s}$-prescription~\cite{Read:2002hq}.

\section{Systematic uncertainties}
\label{sec:systematics}
 
Signal and background expectations are subject to statistical, detector-related and theoretical uncertainties, which are all included in the likelihood as nuisance parameters.
Detector-related and theoretical uncertainties may affect the overall normalisation and/or shape of  the simulated background and signal event distributions.
 
Detector-related uncertainties are dominated by contributions from the jet reconstruction.
Uncertainties in the jet energy scale (JES) for small-$R$ jets~\cite{Aad:2020flx} arise from the calibration of the scale of the jet and are derived as function of the jet \pt, and also $\eta$. Further contributions emerge from the jet flavour composition and the pile-up conditions. The `category reduction' scheme as described in Ref.~\cite{Aad:2020flx} with 29 nuisance parameters is used. Uncertainties in the jet energy resolution (JER) depend on the jet \pt\ and $\eta$ and arise both from the method used to derive the jet resolution and from the difference between simulation and data~\cite{Aad:2020flx}, and they are included with eight nuisance parameters. Similarly, uncertainties in the jet energy resolution for large-$R$ jets arise from the calibration, the flavour composition and the topology dependence~\cite{Aaboud:2018kfi}. Further uncertainties are considered for the large-$R$ jet mass scale~\cite{Aaboud:2018kfi} and resolution~\cite{ATLAS-CONF-2017-063}.
 
Uncertainties due to the $b$-tagging efficiency for heavy-flavour jets, including $c$-flavour jets, are derived from \ttbar\ data~\cite{FTAG-2018-01,ATLAS-CONF-2018-001} and are represented by four nuisance parameters. Uncertainties are also considered for mistakenly $b$-tagging a light-flavour jet, with nine nuisance parameters. These are estimated using a method similar to that in Ref.~\cite{ATLAS-CONF-2018-006}.
 
Uncertainties in the modelling of \met\ are evaluated by considering the uncertainties affecting the jets included in the calculation and the uncertainties in soft term's scale and resolution~\cite{PERF-2016-07}. The pile-up in simulation is matched to the conditions in data by a reweighting factor. An uncertainty of 4\% is assigned to this reweighting factor.
The uncertainty in the combined 2015--2018 integrated luminosity is 1.7\% \cite{ATLAS-CONF-2019-021}, obtained using the LUCID-2 detector \cite{Avoni_2018} for the primary luminosity measurements.
 
Scale factors, including their uncertainties, are calculated specifically for this analysis to correct the efficiency of \met\ triggers in simulation to that in data. The uncertainties in the scale factors are at most 1\%--2\% for low \met\ values.
 
In the regions requiring the presence of leptons, uncertainties in the lepton identification and lepton energy/momentum scale and resolution are included. These are derived using simulated and measured events with $Z \rightarrow \ell^{+}\ell^{-}$, $J/\psi \rightarrow \ell^{+}\ell^{-}$ and $W \rightarrow \ell\nu$ decays~\cite{EGAM-2018-01,MUON-2018-03}.
 
Modelling uncertainties impact the shape of the $m_h$ distribution, the relative acceptance between different \met\ and $b$-tag multiplicity bins and between signal and control regions as well as the overall normalisation of the samples that are not freely floating in the fit.
 
The theoretical uncertainties are dominated by modelling uncertainties in the \ttbar\  and $Z$+HF backgrounds.
For the \ttbar and $Wt$ processes, the impact of the choice of parton shower and hadronisation model is evaluated by comparing the sample from the nominal generator set-up
with a sample interfaced to \herwigseven.04~\cite{Bahr:2008pv,Bellm:2015jjp}.
To assess the uncertainty in the matching of NLO matrix elements to the parton shower, the \powhegbox\ sample is compared with a sample of events generated with \mgamc~v2.6.2. 
For the $Wt$ process, the nominal sample is compared with an alternative sample generated using the diagram subtraction scheme~\cite{Frixione:2008yi,ATL-PHYS-PUB-2016-020} instead of the diagram removal scheme to estimate the uncertainty arising from the interference with \ttbar\ production.
 
For the $V$+jet processes, uncertainties arising from the modelling of the parton shower and the matching scheme are evaluated by comparing the nominal samples with samples generated with \mgamc~v2.2.2.
For the diboson processes, the uncertainties associated with the modelling of the parton shower, the hadronisation and the underlying event are derived using alternative samples generated with \powhegbox~\cite{Nason:2004rx,Frixione:2007vw,Alioli:2010xd} and interfaced to \pythia.186~\cite{Sjostrand:2007gs} or \textsc{Herwig++}.
 
For all MC samples, the uncertainties due to missing higher orders are estimated by a variation of the renormalisation and factorisation scales by a factor of two, while the PDF and \alphas uncertainties are calculated using the PDF4LHC prescription~\cite{Butterworth:2015oua}.
 
Table~\ref{syst:table:impact} gives the impact of the different sources of systematic uncertainties for selected signal models as evaluated in different model-dependent fits. The signal models with lower masses illustrate the impact of the systematic uncertainties in the resolved regions, while the models with larger mediator masses are more impacted by the merged regions. The theoretical uncertainties in the modelling of the \ttbar\ background, the experimental uncertainties in the calibration of jets and the limited MC sample size show the largest impact.
 
\begin{table}
\caption{\label{syst:table:impact} Relative importance of the different sources of uncertainty for different $Z'$-2HDMs, with the masses of the $Z'$ boson and the $A$ boson given in the second row, expressed as fractional impact on the signal strength parameter. The fractional impact is calculated by considering the square of the uncertainty in the signal strength parameter arising from a given group of uncertainties (as listed in the left column of the table), divided by the square of the total uncertainty in the signal strength parameter. Due to correlations, the sum of the different impacts of systematic uncertainties might not add up to the total impact of all systematic uncertainties.}
\vspace{0.2cm}
\begin{center}
\begin{tabular}{l|ccc}
\hline
\multirow{3}{*}{Source of uncertainty} & \multicolumn{3}{c}{Fractional squared uncertainty in $\mu$} \\\cline{2-4}
& \multicolumn{3}{c}{$Z'$-2HDM signals, $(m_Z',m_A)$ [\GeV]} \\
& (800, 500) & (1400, 1000) & (2800, 300) \\
\hline
$Z$+HF normalisation & 0.11 & 0.03 & <0.01~~~\\
$W$+HF normalisation & 0.02 & 0.01 & <0.01~~~\\
\ttbar\ normalisation & 0.16 & 0.04 & <0.01~~~\\
$Z$ modelling uncertainties & 0.02 & 0.07 & <0.01~~~\\
$W$ modelling uncertainties & <0.01~~~& 0.01 & <0.01~~~\\
\ttbar\ modelling uncertainties & 0.13 & 0.05 & <0.01~~~\\
Single-$t$ modelling uncertainties& 0.18 & 0.02 & <0.01~~~\\
Other modelling uncertainties & 0.05 & 0.01 & <0.01~~~\\
Jets & 0.20 & 0.06 & 0.01 \\
$b$-tagging & 0.01 & 0.01 & 0.04 \\
\met\ soft term and pile-up & <0.01~~~& <0.01~~~& <0.01~~~\\
Other experimental systematic uncertainties & 0.01 & <0.01~~~& <0.01~~~\\
Signal systematic uncertainties & <0.01~~~& <0.01~~~& <0.01~~~\\
MC sample size & 0.08 & 0.07 & 0.11 \\
\hline
Statistical uncertainty & 0.27 & 0.61 & 0.79 \\
Total systematic uncertainties & 0.73 & 0.39 & 0.21 \\\hline
\end{tabular}
\end{center}
\end{table}
 
\section{Results}
\label{sec:results}
 
\begin{figure}[p]
\centering
\begin{subfigure}[b]{\textwidth}
\includegraphics[width=\textwidth]{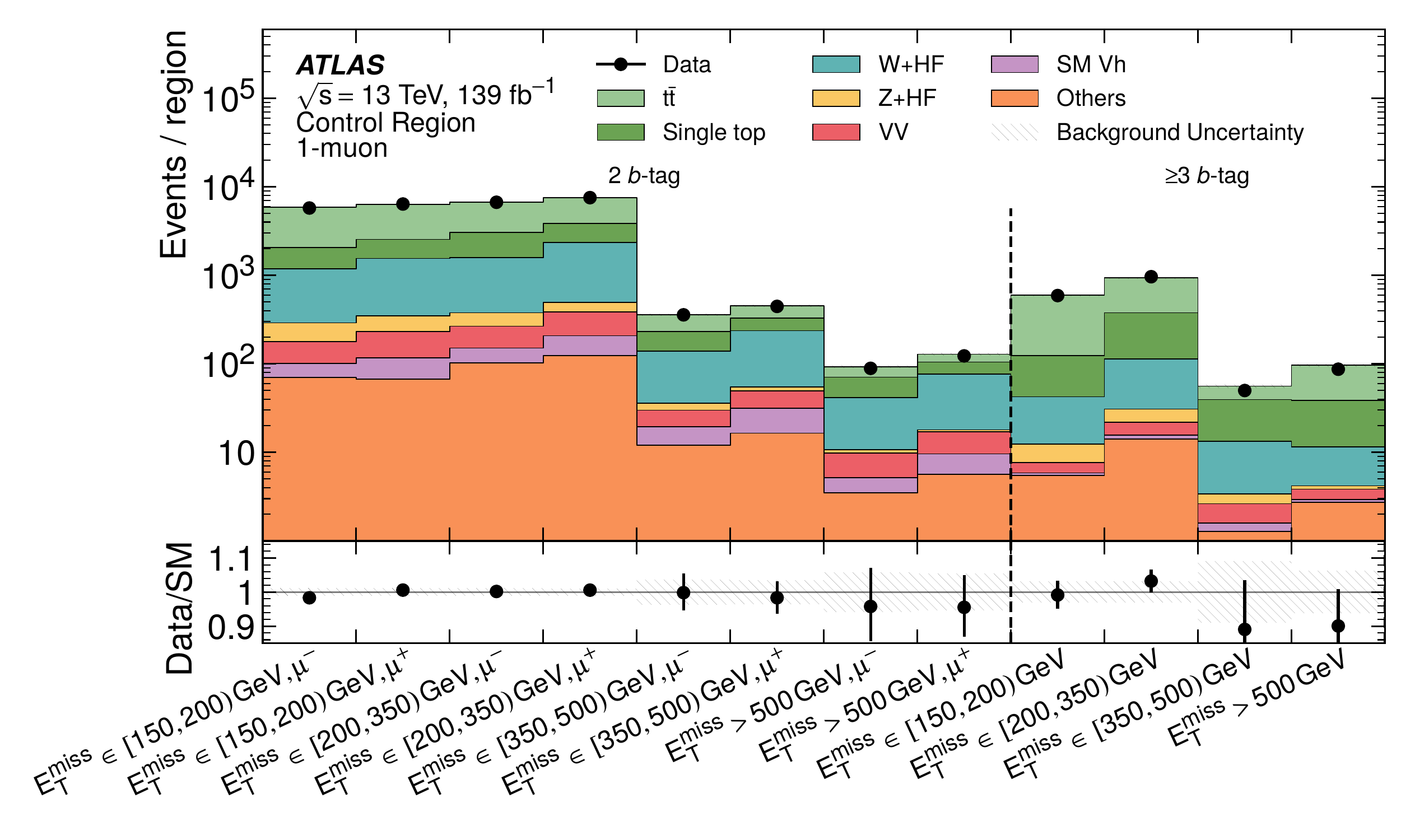}
\caption{\label{fit:fig:CR.CR1}}
\end{subfigure}
\begin{subfigure}[b]{0.7\textwidth}
\includegraphics[width=\textwidth]{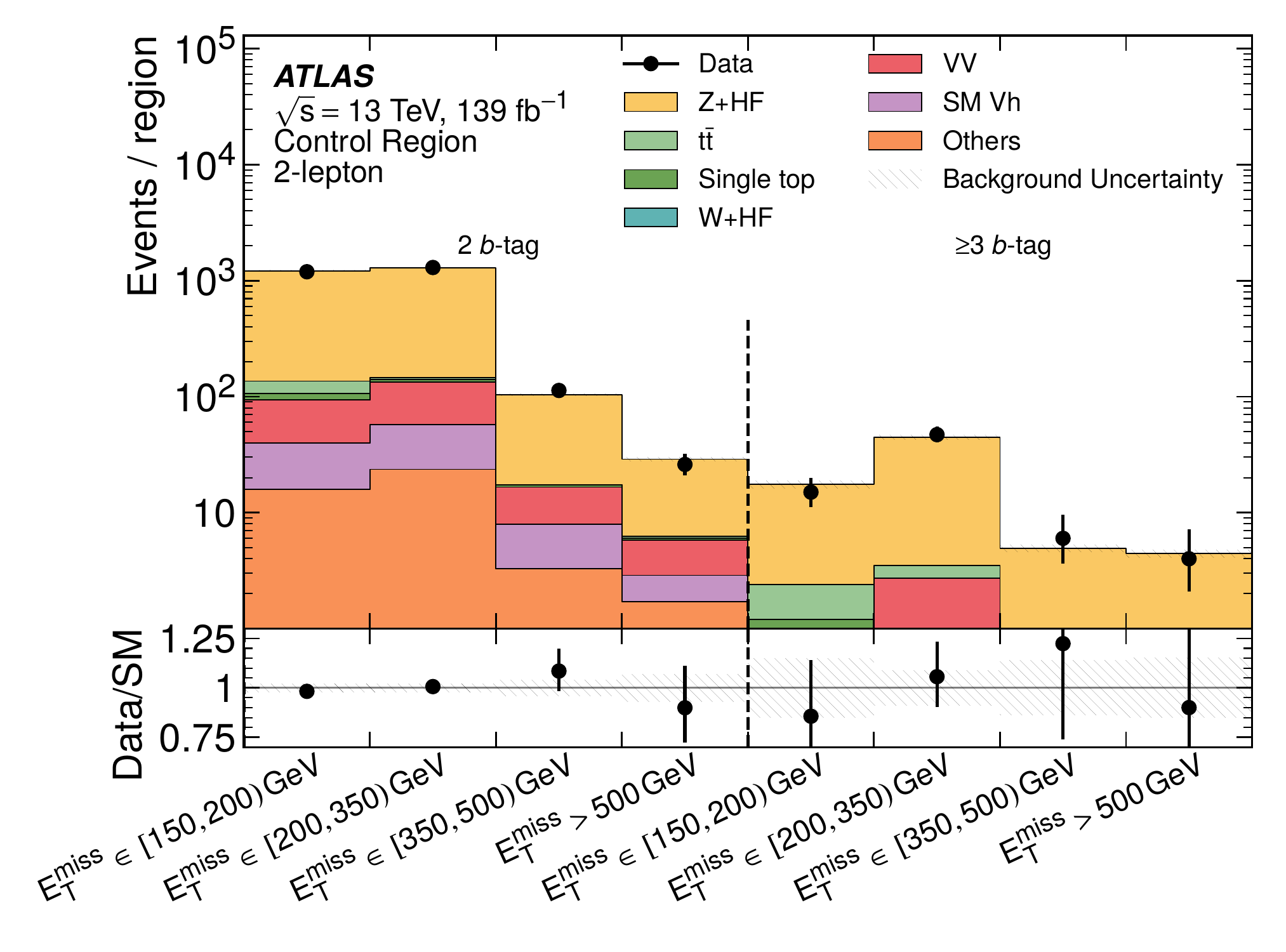}
\caption{\label{fit:fig:CR.CR2}}
\end{subfigure}
\caption{\label{fit:fig:CR}Yields in the resolved and merged \subref{fit:fig:CR.CR1}~1-muon control regions and \subref{fit:fig:CR.CR2}~2-lepton control regions. The top panel  compares the fitted background yields with data, while the bottom panel indicates the ratio of the observed data to the predicted Standard Model backgrounds. The different control region bins included in the fit are indicated on the $x$-axis by first giving the range in \MET\ and then the sign of the muon charge (where applicable).}
\end{figure}

The post-fit background yields are determined in a background-only profile likelihood fit to data in all regions.
Figure~\ref{fit:fig:CR} shows the yields in the 1-muon and 2-lepton control regions.
The post-fit normalisation factors for \ttbar\ and for $W$+HF are found to be $0.93 \pm 0.08$ and $0.95 \pm 0.14$, respectively. For $Z$ boson production in association with two (at least three) heavy-flavour jets the normalisation factors are determined to $1.41 \pm 0.09$ ($1.85 \pm 0.24$). An upward scaling of the $Z$+HF background relative to the simulation was also observed in other studies~\cite{STDM-2017-38}, and was attributed to an underestimation of the $g \rightarrow b\bar{b}$ rate in \textsc{Sherpa}. A larger scaling is observed in the region with ${\geq}3$ $b$-tagged jets, dominated by processes with more $g \rightarrow b\bar{b}$ splittings. The uncertainty on the $Z$+HF normalisation factor increases in the \3b\ region due to lower statistical precision and the smaller contribution of the $Z$+HF background.
 
The distributions of the Higgs boson candidate mass $m_{h}$ after the background-only fit are shown in Figures~\ref{results:fig:SRplots1} and \ref{results:fig:SRplots2}. The signal to background ratio is higher in the \3b\ signal regions, because the 2HDM+$a$ signal model is shown with $\tan\beta=10$, where $b$-associated
production dominates. Tables~\ref{results:tables:SRyields1} and~\ref{results:tables:SRyields2} present the background estimates in comparison with the observed data. No significant deviation from SM expectations is observed, with the largest deficit corresponding to a local significance of $2.3 \sigma$, and the largest excess amounting to $1.6 \sigma$. Figure~\ref{fig:results:metdist} summarises the total yields in the signal regions as a function of \MET. The background prediction from simulation is scaled upwards in the fit for lower \MET\ values, while the simulation agrees better with the data for large \MET\ values.
 
\begin{table}
\centering
\caption{\label{results:tables:SRyields1}Background yields in comparison with data in the \2b\ signal regions for different \met\ ranges after a background-only fit to data. Statistical and systematic uncertainties are reported together.}
\vspace{0.2cm}
\footnotesize
\begin{tabular}{l|c|c|c|c|c}
\hline
\2b\ signal regions & \multirow{2}{*}{$[150,200)~\GeV$} & \multirow{2}{*}{$[200,350)~\GeV$} & \multirow{2}{*}{$[350,500)~\GeV$} & \multirow{2}{*}{$ [500,750)~\GeV$} & \multirow{2}{*}{$ > 750~\GeV$}\\
\MET\ range &  &   &  &   &  \\
\hline
$Z$+HF & 6470 $\pm$ 310 & 7200 $\pm$ 310 & 507 $\pm$ 26 & 94 $\pm$ 7 & 9.2 $\pm$ 1.8\\
$Z$+light jets & ~~72 $\pm$ 15 & 137 $\pm$ 29 & 18 $\pm$ 4 & ~~4.5 $\pm$ 1.0 & 1.17 $\pm$ 0.30\\
$W$+HF & 1590 $\pm$ 210 & 1760 $\pm$ 230 & 106 $\pm$ 14 & 25 $\pm$ 4 & 3.1 $\pm$ 0.6\\
$W$+light jets & ~~86 $\pm$ 35 & ~~92 $\pm$ 35 & 14 $\pm$ 5 & ~~1.6 $\pm$ 0.6 & 0.21 $\pm$ 0.09\\
Single top-quark & ~~570 $\pm$ 260 & ~~570 $\pm$ 260 & ~~21 $\pm$ 10 & ~~2.6 $\pm$ 1.9 & 0.10 $\pm$ 0.16\\
\ttbar\ & 4680 $\pm$ 290 & 3280 $\pm$ 240 & 76 $\pm$ 9 & 11.4 $\pm$ 1.6 & 0.38 $\pm$ 0.08\\
Diboson & 450 $\pm$ 50 & 600 $\pm$ 60 & 56 $\pm$ 7 & 15.2 $\pm$ 1.9 & 1.61 $\pm$ 0.29\\
$Vh$ & 151 $\pm$ 10 & 202 $\pm$ 12 & 26.6 $\pm$ 1.8 & ~~5.6 $\pm$ 0.5 & 0.68 $\pm$ 0.12\\
\ttbar+$V/h$ & ~~7.6 $\pm$ 0.4 & 11.8 $\pm$ 0.5 & ~~0.45 $\pm$ 0.06 & ~~0.286 $\pm$ 0.029 & 0.035 $\pm$ 0.006\\
\hline
Total background & 14070 $\pm$ 110~~ & 13860 $\pm$ 100~~ & 825 $\pm$ 19 & 160 $\pm$ 8~~ & 16.7 $\pm$ 1.9~~ \\
\hline
Data & 14259 & 13724 & 799 & 168 & 19\\ \hline
\end{tabular}
\end{table}

\begin{table}
\centering
\caption{\label{results:tables:SRyields2}Background yields in comparison with data in the \3b\ signal regions for different \met\ ranges after a background-only fit to data. Statistical and systematic uncertainties are reported together.}
\vspace{0.2cm}
\small
\begin{tabular}{l|c|c|c|c}
\hline
\3b\ signal regions & \multirow{2}{*}{$[150,200)~\GeV$} & \multirow{2}{*}{$[200,350)~\GeV$} & \multirow{2}{*}{$[350,500)~\GeV$} & \multirow{2}{*}{$ > 500~\GeV$}\\
\MET\ range & & & &\\
\hline
$Z$+HF & 102 $\pm$ 15~~ & 278 $\pm$ 28~~ & 26.4 $\pm$ 3.5~~ & 15.6 $\pm$ 1.9~~ \\
$Z$+light jets & 0.6 $\pm$ 0.4 & 2.9 $\pm$ 0.8 & 0.34 $\pm$ 0.12 & 0.46 $\pm$ 0.12\\
$W$+HF & 21 $\pm$ 4~~ & 47 $\pm$ 9~~ & 4.2 $\pm$ 0.9 & 2.4 $\pm$ 0.4\\
$W$+light jets & 0.01 $\pm$ 0.04 & 1.7 $\pm$ 0.9 & 0.8 $\pm$ 0.4 & 0.031 $\pm$ 0.026\\
\ttbar\ & 276 $\pm$ 19~~ & 252 $\pm$ 22~~ & 5.1 $\pm$ 0.7 & 17.9 $\pm$ 1.8~~ \\
Single top-quark & 23 $\pm$ 11 & 55 $\pm$ 25 & 2.9 $\pm$ 1.4 & 3.4 $\pm$ 1.7\\
Diboson & 4.8 $\pm$ 1.4 & 12.9 $\pm$ 2.2~~ & 1.8 $\pm$ 0.4 & 1.26 $\pm$ 0.31\\
$Vh$ & 0.65 $\pm$ 0.28 & 2.9 $\pm$ 0.5 & 0.40 $\pm$ 0.08 & 0.230 $\pm$ 0.025\\
\ttbar+$V/h$ & 1.78 $\pm$ 0.17 & 3.89 $\pm$ 0.26 & 0.371 $\pm$ 0.035 & 0.78 $\pm$ 0.08\\
\hline
Total background & 430 $\pm$ 15~~ & 656 $\pm$ 21~~ & 42 $\pm$ 4~~ & 42.0 $\pm$ 2.8~~ \\
\hline
Data & 408 & 658 & 42 & 46\\ \hline
\end{tabular}
\end{table}
 
\begin{figure}[tbh]
\centering
\begin{subfigure}[b]{0.46\textwidth}
\includegraphics[width=\textwidth]{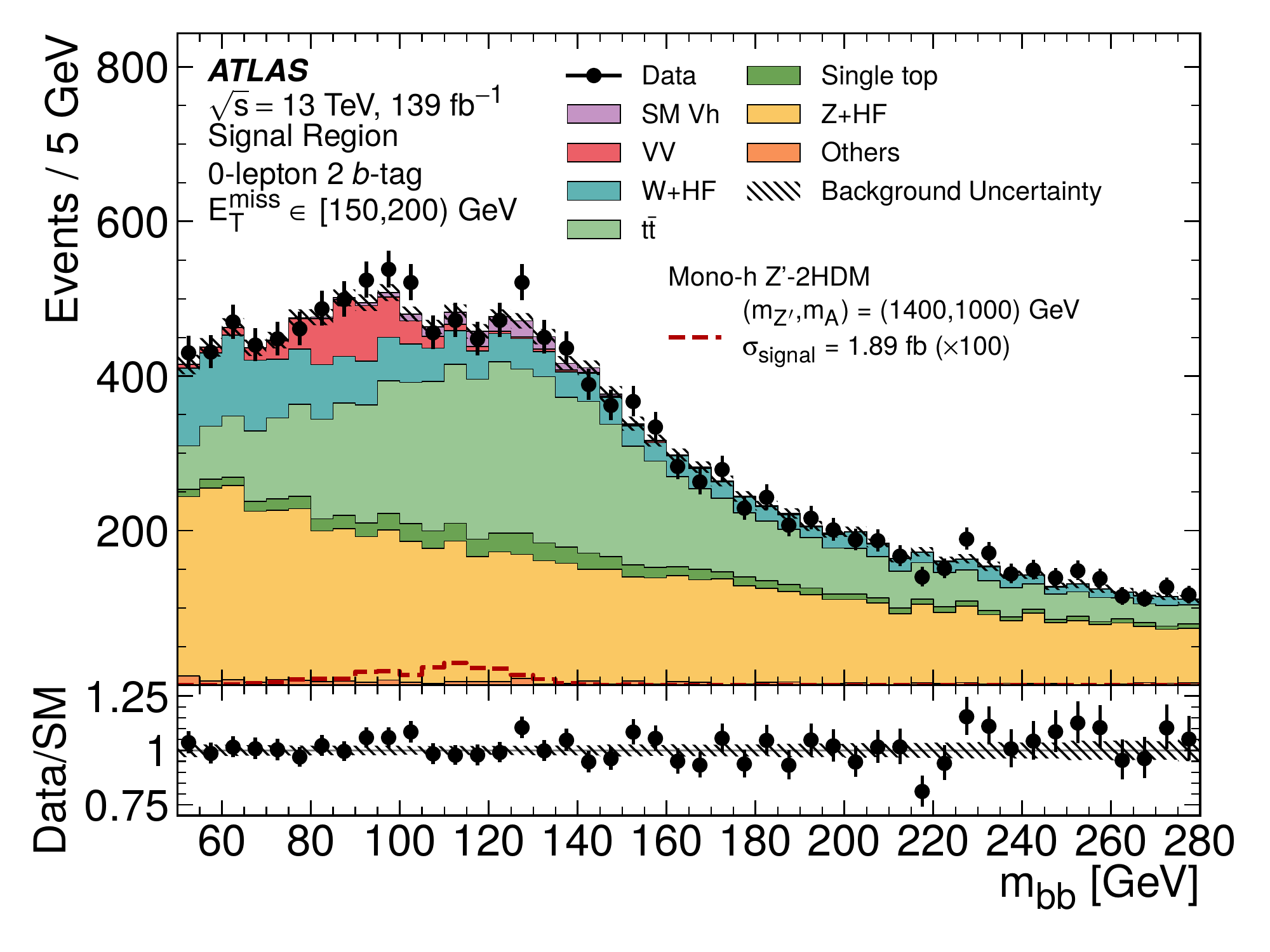}
\caption{}
\end{subfigure}
\begin{subfigure}[b]{0.46\textwidth}
\includegraphics[width=\textwidth]{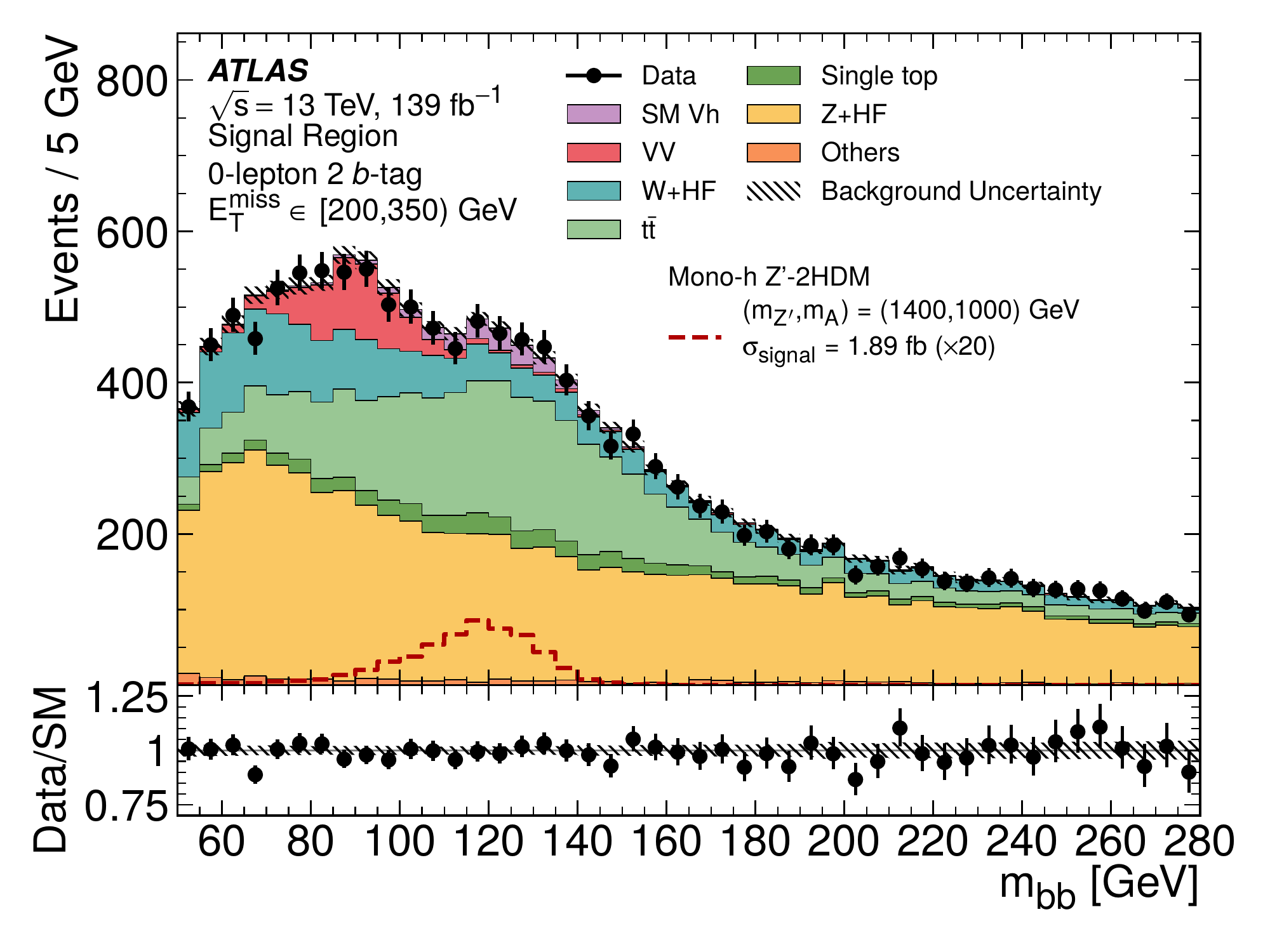}
\caption{}
\end{subfigure}
\begin{subfigure}[b]{0.46\textwidth}
\includegraphics[width=\textwidth]{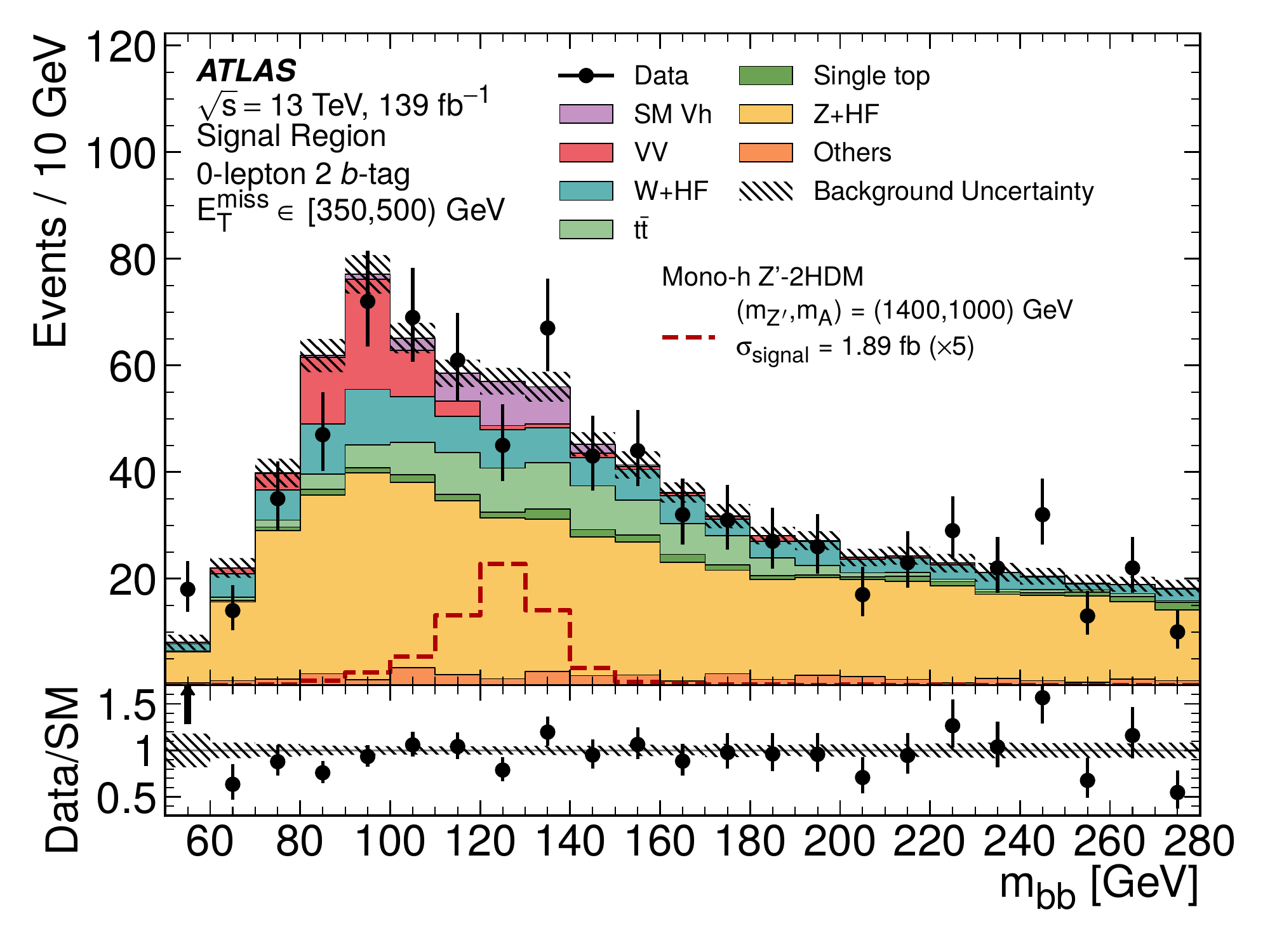}
\caption{}
\end{subfigure}
\begin{subfigure}[b]{0.46\textwidth}
\includegraphics[width=\textwidth]{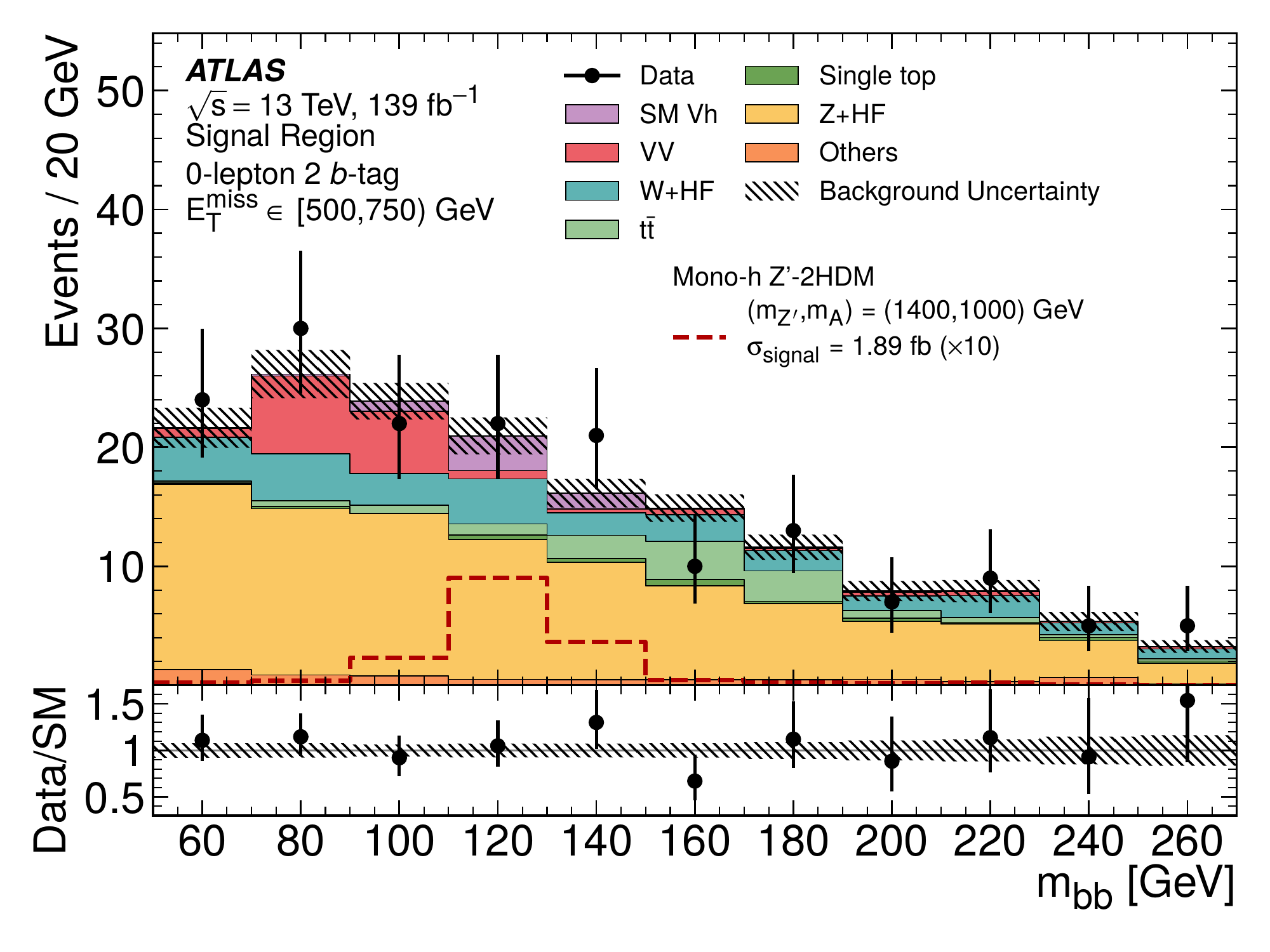}
\caption{}
\end{subfigure}
\begin{subfigure}[b]{0.46\textwidth}
\includegraphics[width=\textwidth]{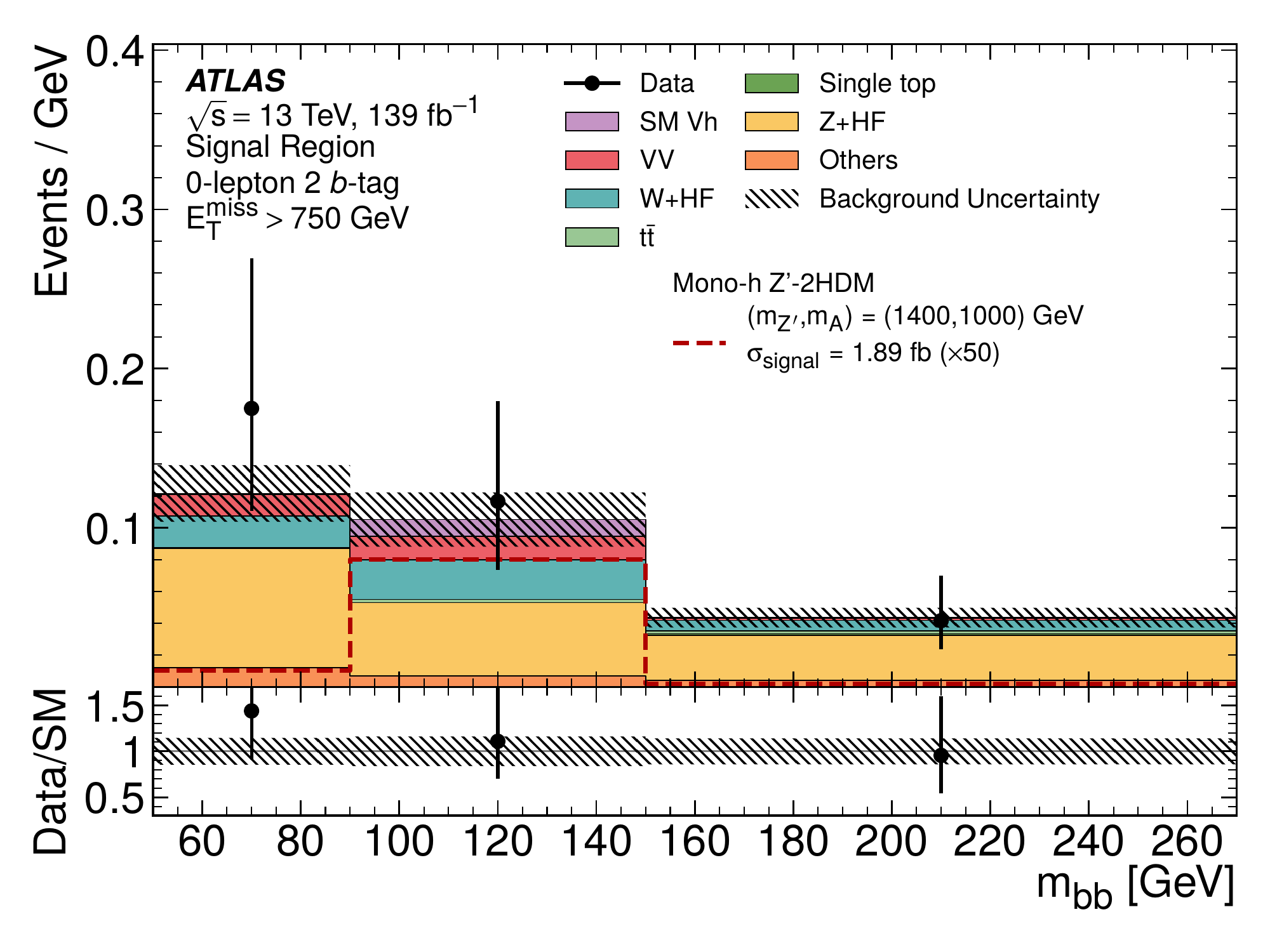}
\caption{}
\end{subfigure}
\caption{\label{results:fig:SRplots1}Distributions of the Higgs boson candidate mass in the \2b\ signal regions for different \met\ ranges. The top panel comparies the fitted background yields with data, while the bottom panel indicates the ratio of the observed data to the predicted Standard Model backgrounds. The background yields are obtained in a background-only fit to data. An example signal model from the $Z'$-2HDM with parameters $(m_{Z'},m_{A}) = (1400~\GeV,1000~\GeV)$ and a cross-section of 1.89~fb is displayed for comparison. The signal is scaled by the factors indicated in the legend for better visibility.}
\end{figure}
 
\clearpage
 
\begin{figure}[tbh]
\begin{subfigure}[b]{0.49\textwidth}
\includegraphics[width=\textwidth]{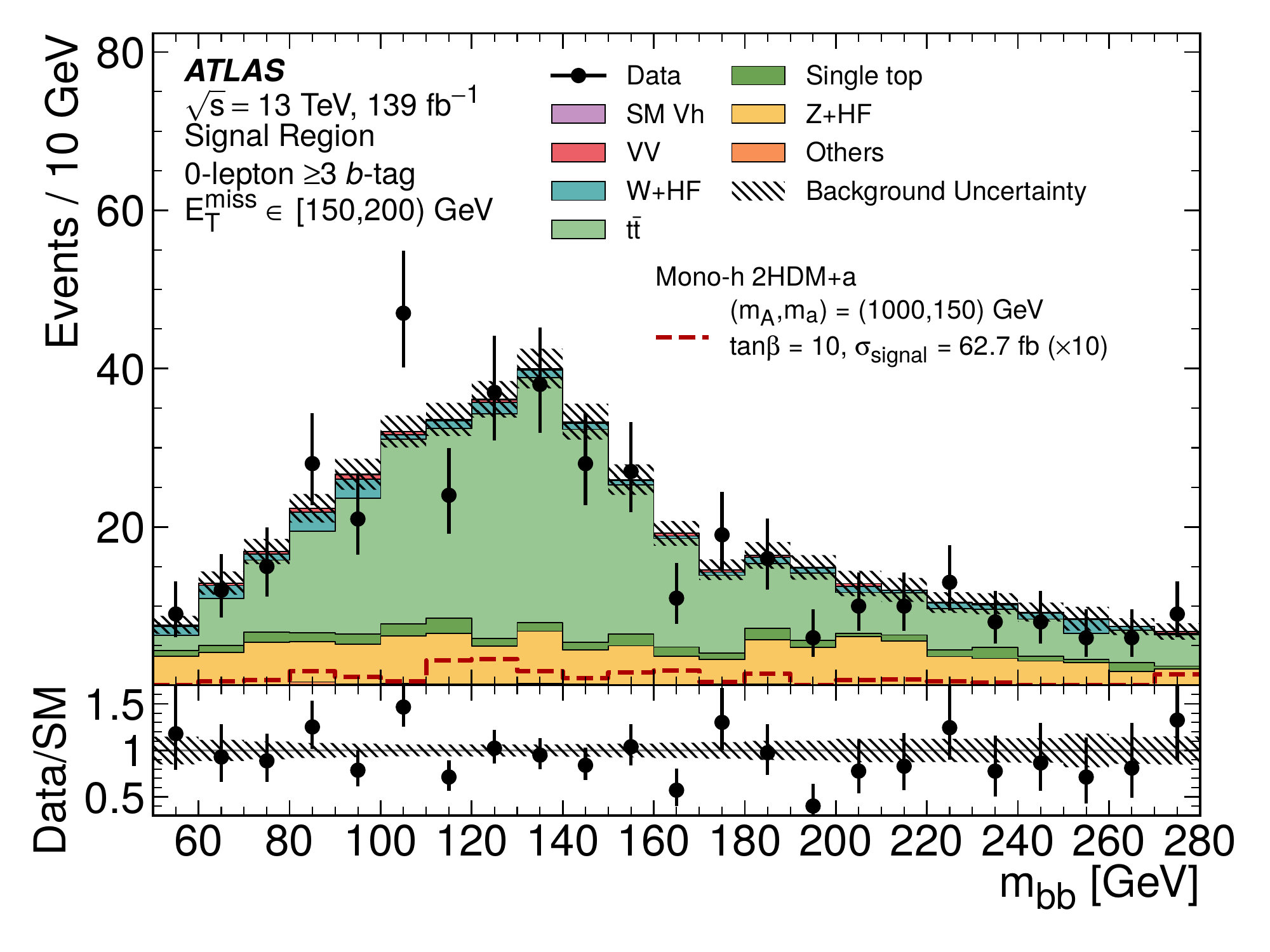}
\caption{}
\end{subfigure}
\begin{subfigure}[b]{0.49\textwidth}
\includegraphics[width=\textwidth]{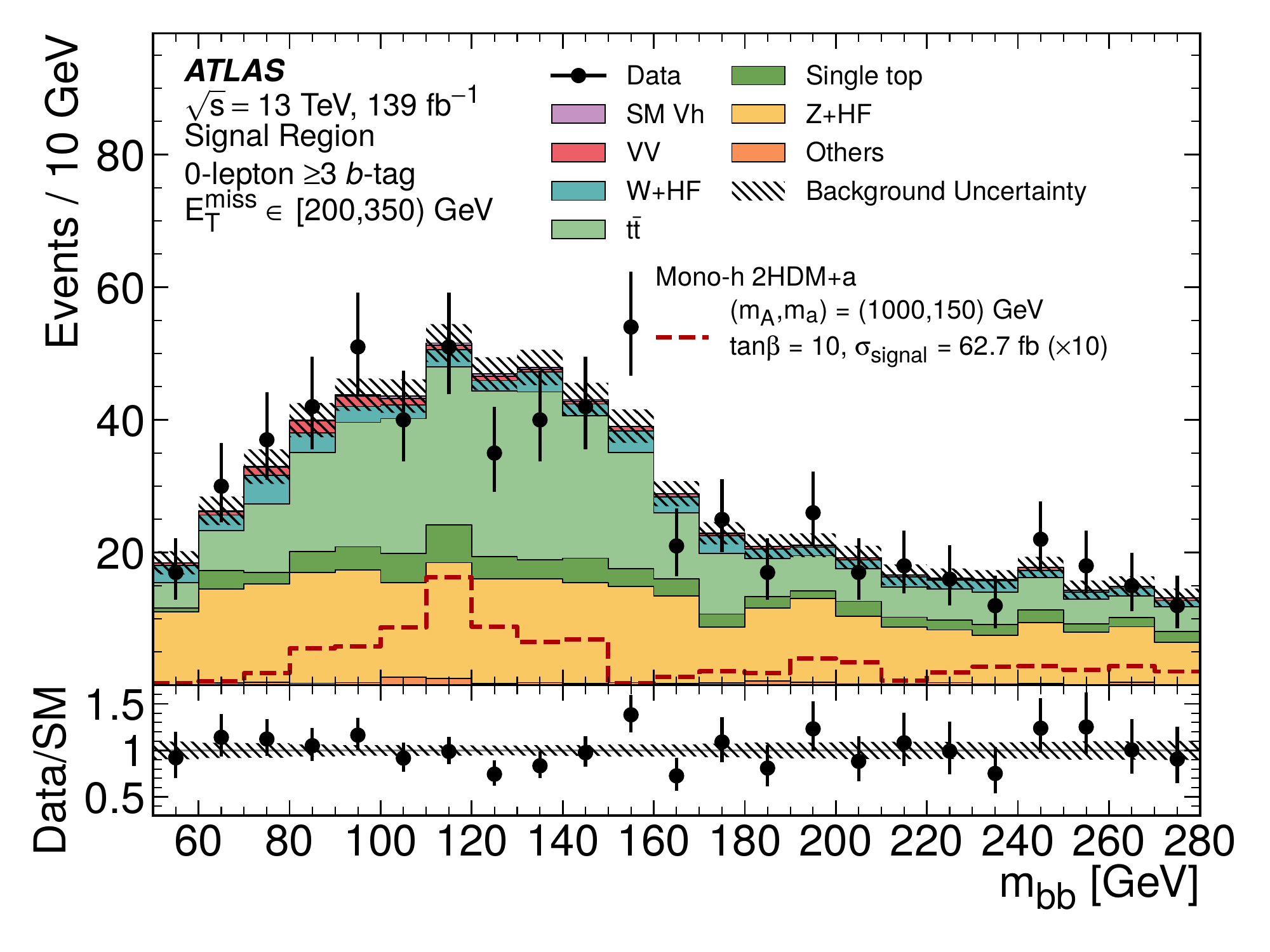}
\caption{}
\end{subfigure}
\begin{subfigure}[b]{0.49\textwidth}
\includegraphics[width=\textwidth]{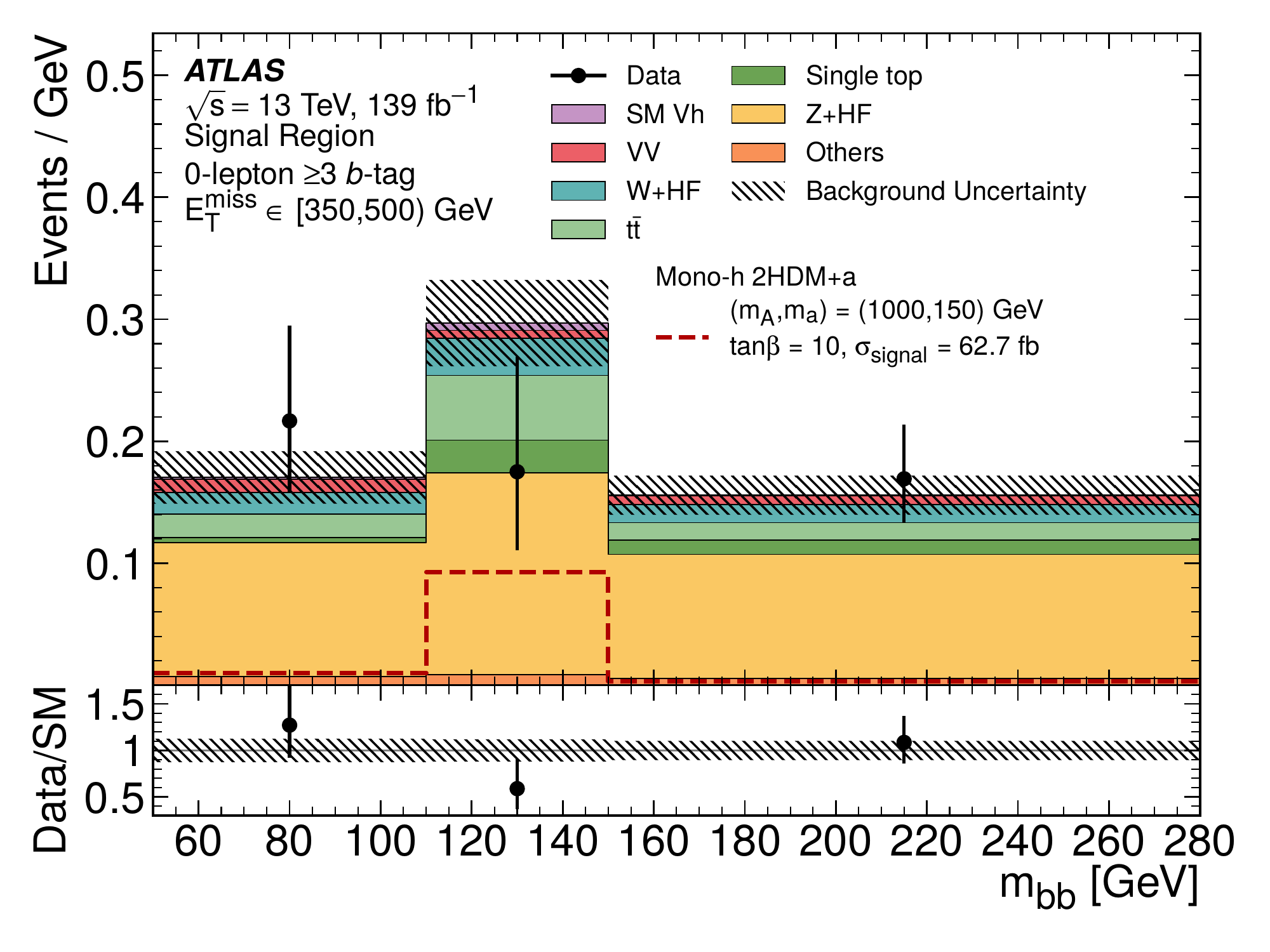}
\caption{}
\end{subfigure}
\begin{subfigure}[b]{0.49\textwidth}
\includegraphics[width=\textwidth]{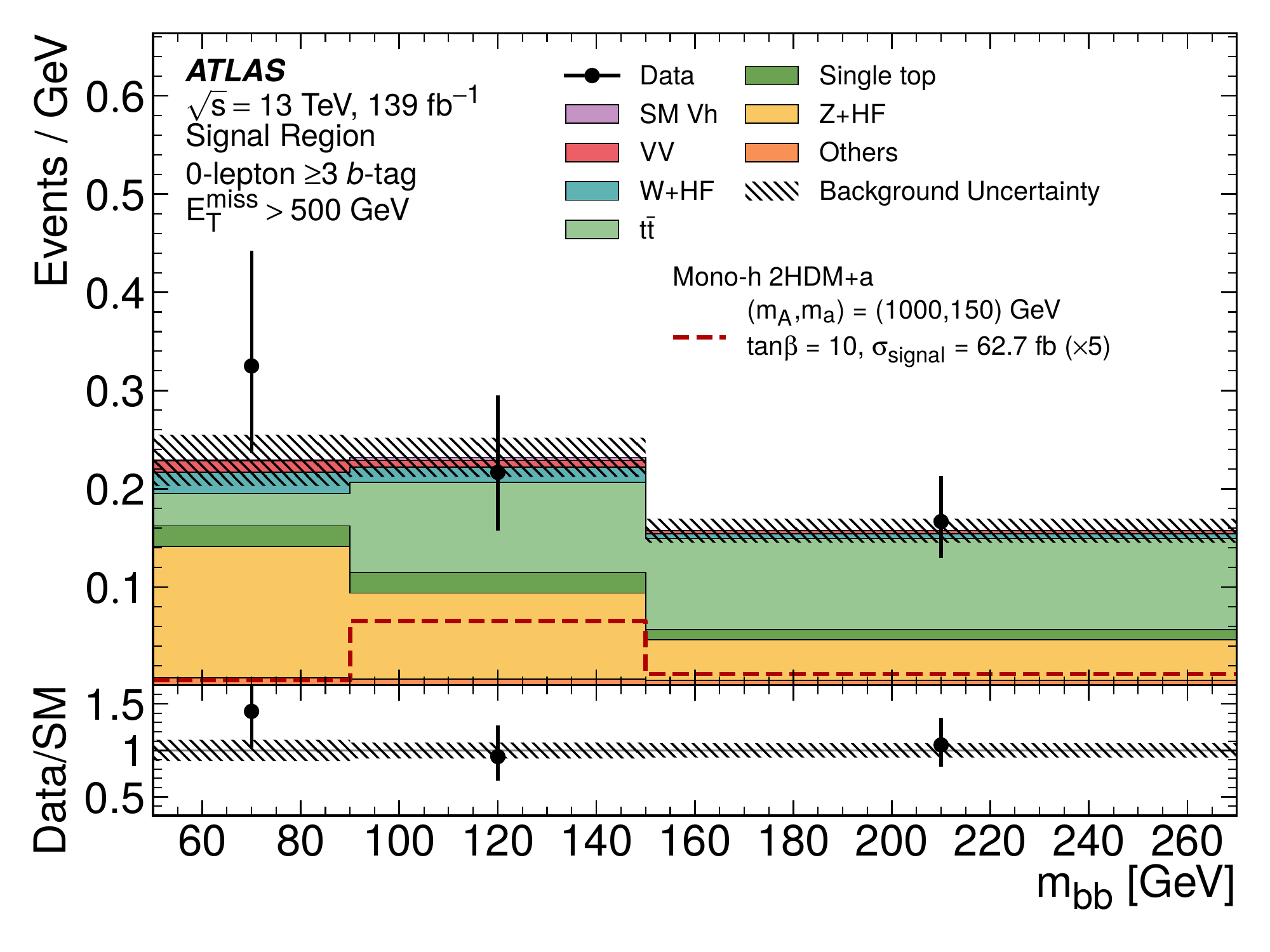}
\caption{}
\end{subfigure}
\caption{\label{results:fig:SRplots2}Distributions of the Higgs boson candidate mass in the \3b\ signal regions. The top panel compares the fitted background yields with data, while the bottom panel indicates the ratio of the observed data to the predicted Standard Model backgrounds. The background yields are obtained in a background-only fit to data. An example signal model from the 2HDM+$a$ with $bbA$ production and with parameters $(m_{A},m_{a}) = (1000~\GeV,150~\GeV)$, $\tan \beta = 10$, and a cross-section of 62.7~fb is displayed for comparison. The signal is scaled by the factors indicated in the legend for better visibility.}
\end{figure}

\begin{figure}[tbh]
\begin{subfigure}[b]{0.49\textwidth}
\includegraphics[width=\textwidth]{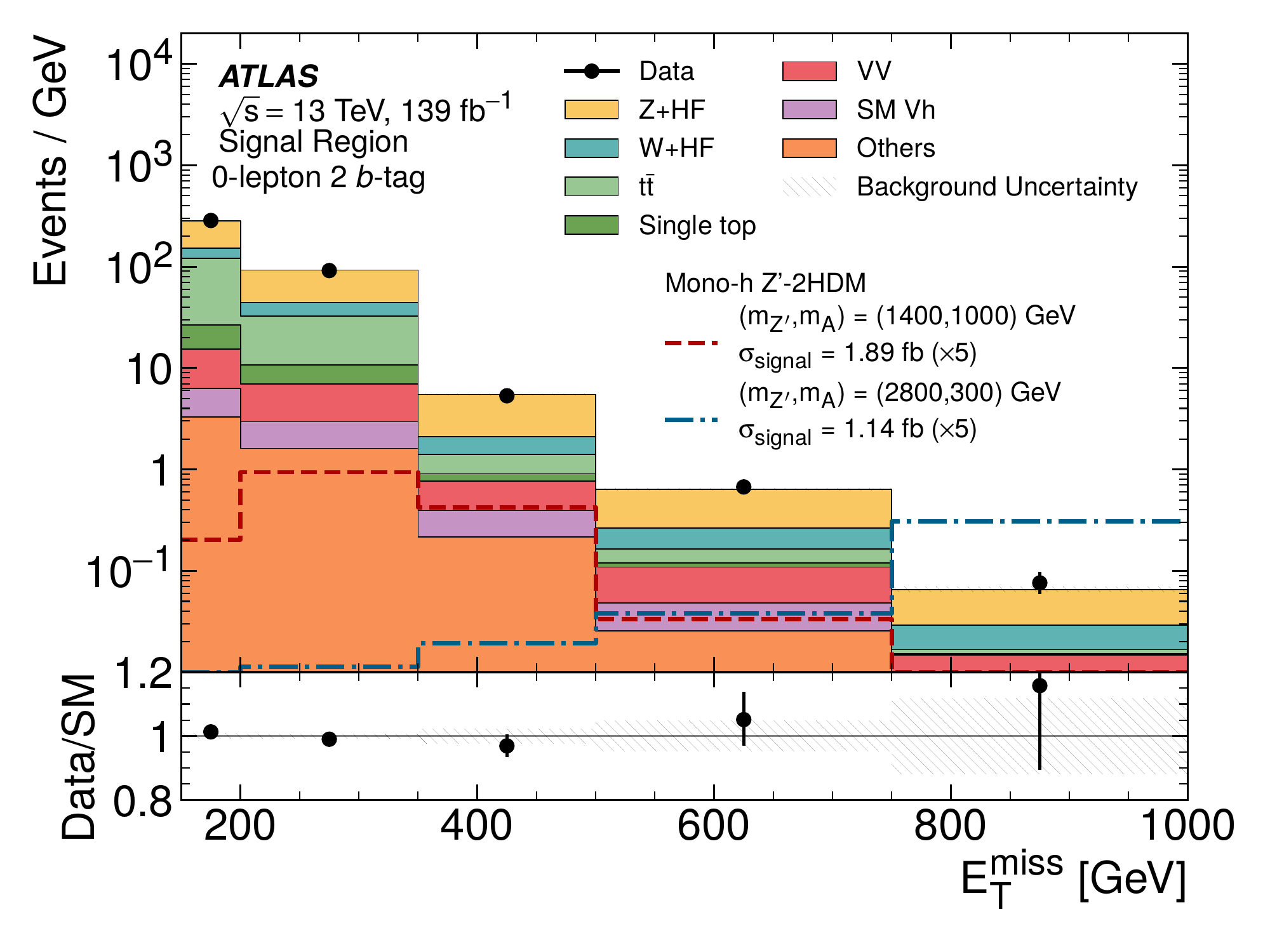}
\caption{\label{fig:results:metdist.2b}}
\end{subfigure}
\begin{subfigure}[b]{0.49\textwidth}
\includegraphics[width=\textwidth]{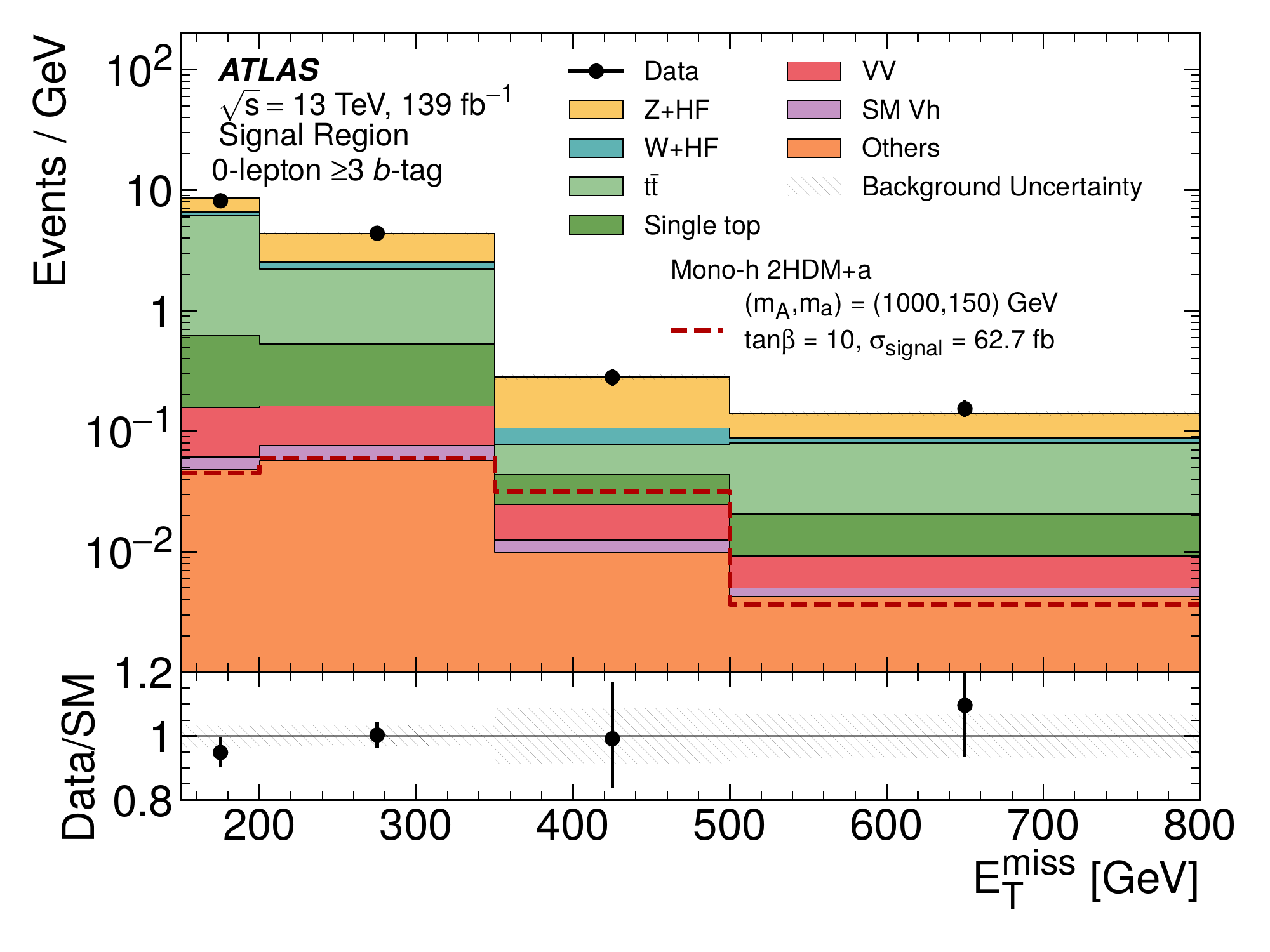}
\caption{\label{fig:results.metdist.3b}}
\end{subfigure}
\caption{\label{fig:results:metdist}\MET\ distributions after requiring \subref{fig:results:metdist.2b}~two $b$-tagged jets or \subref{fig:results.metdist.3b}~at least three $b$-tagged jets.
The background yields are obtained in a background-only fit to data.
Each bin corresponds to one signal region, where the bins at lower \MET\ use small-$R$ jets and the ones with larger \MET\ select events with at least one large-$R$ jet.
The rightmost bin includes all events with \MET\ above the range shown on the plots.
The bottom panel compares the data with background estimates as obtained from the fit.
Example signal models are overlaid in the top panel.
In plot~\subref{fig:results:metdist.2b} two $Z'$-2HDM points are shown, one with parameters $(m_{Z'},m_{A}) = (1400~\GeV,1000~\GeV)$ and a cross-section of 1.89~fb and the other with parameters $(m_{Z'},m_{A}) = (2800~\GeV,300~\GeV)$ and a cross-section of 1.14~fb.
Both signals are scaled up by a factor of 5 for better visibility.
In plot~\subref{fig:results.metdist.3b} a 2HDM+$a$ is shown with parameters $(m_{A},m_{a}) = (1000~\GeV,150~\GeV)$, $\tan \beta = 10$ and a cross-section of 62.7~fb.}
\end{figure}
 
The results are interpreted as exclusion limits at 95\% CL in the $Z'$-2HDM and the 2HDM+$a$ scenarios
in Figures~\ref{results:fig:contourplots2} and \ref{results:fig:contourplots}. Considering the $Z'$-2HDM case, $Z'$ masses up to 3 \TeV\ are excluded for $A$ masses of 300~\GeV\ at 95\% CL. The exclusion boundaries for the 2HDM+$a$ scenario extend up to $m_a = 520~\GeV$ for $m_A = 1.25~\TeV$ for $gg$F production and $\tan \beta = 1$. This is an improvement of about 200~\GeV\ in $m_a$ on previous results~\cite{EXOT-2017-32}, which reinterpreted the earlier $h(\rightarrow b\bar{b})$ + \MET\ analysis using 36.1~\ifb~\cite{EXOT-2016-25}.
 
The higher exclusion limit at high $m_{A}$, low $m_{a}$, is due to an increase of the cross-section of the $a \rightarrow ah$ process, without resonant $A$ production. It should be noted that with the exact parameter choices adopted in this analysis, the $aah$ coupling becomes larger than $4\pi$ for $m_A \gtrsim 1750\,\GeV$. Moreover, as discussed in Refs.~\cite{Abe:2018bpo,Bauer_2017} values of $m_A \gtrsim 1250\,\GeV$ (for $\tan \beta=1$) or $m_A \gtrsim 2150\,\GeV$ (for $\tan \beta=10$) would not be consistent with the requirement of having a bounded-from-below scalar potential, given the parameter choices discussed in this paper. These constraints can be relaxed substantially if the quartic couplings assume a value closer to the perturbativity limit and also in more general 2HDMs containing additional couplings as discussed in Refs.~\cite{Bauer:2017fsw,Haisch:2018djm}. Therefore, the above should not be considered as a strong requirement for the validity of the model predictions. At high $m_{A}$ the width of the additional Higgs bosons grows substantially and the theoretical predictions are subject to additional theoretical uncertainties associated with the treatment of the width. Exclusion limits are therefore not shown in the region of very large widths ($m_A>2200$ GeV).
 
In the case of $bbA$ production and $\tan \beta = 10$, the exclusion limits extend up to $m_a = 240~\GeV$ for $m_A = 900~\GeV$. The inclusion of the \3b\ tag region helps to increase the sensitivity relative to the \2b\ tag region by about 30--70\%. The difference between observed and expected limits arises from data deficits in the \3b\ region, especially the deficit around the Higgs boson peak in the $\MET \in [350,500)~\GeV$ region, as shown in Figure~\ref{results:fig:SRplots2}. The 2HDM+$a$ scenario with $bbA$ production and $\tan \beta = 10$ is considered for the signatures discussed in this paper for the first time.

\begin{figure}
\centering
\begin{subfigure}[b]{0.7\textwidth}
\includegraphics[width=\textwidth]{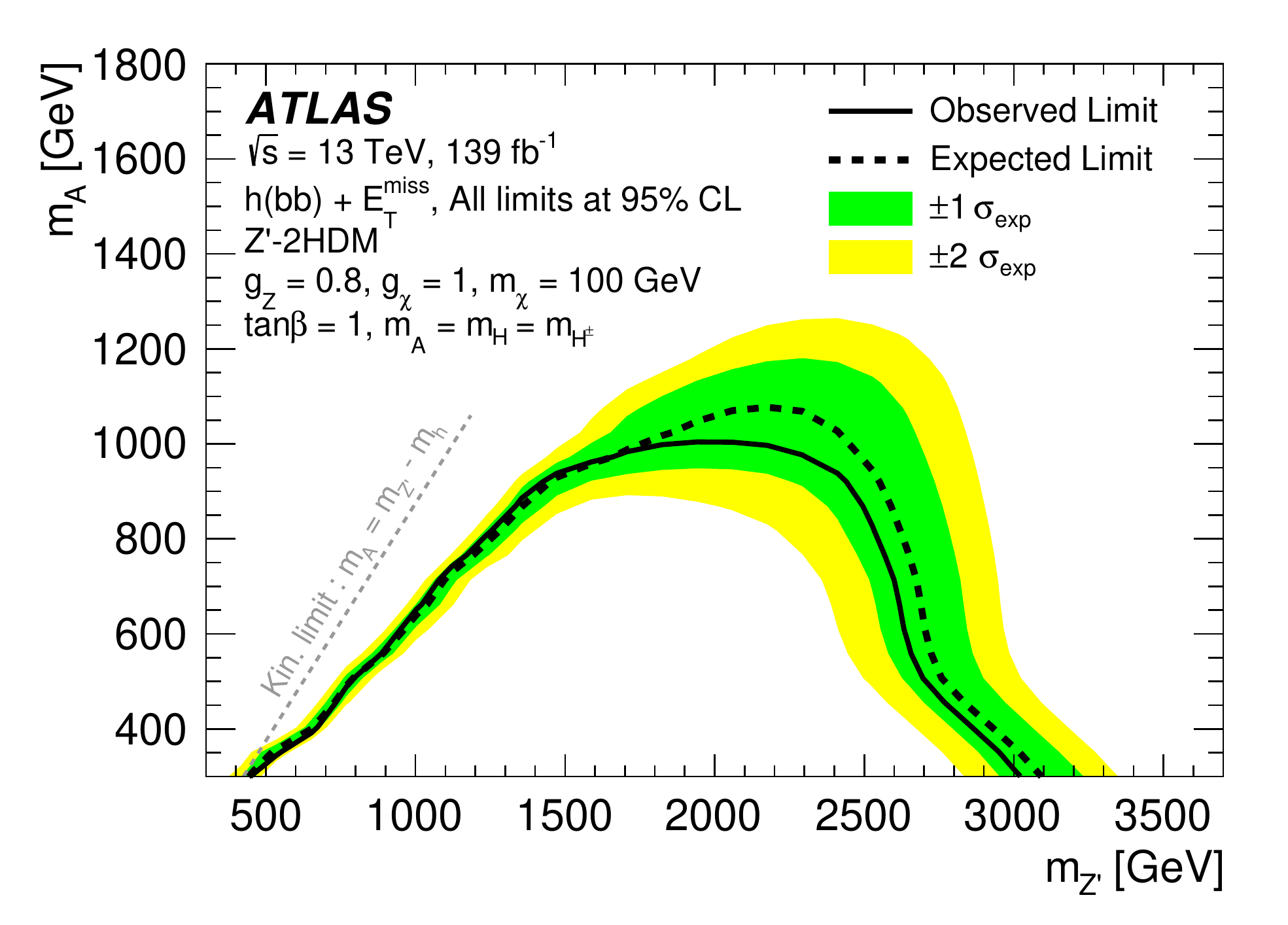}
\end{subfigure}
\caption{\label{results:fig:contourplots2} Exclusion limits in the $Z^{'}$-2HDM model. The solid black line shows the observed limit at 95\% CL, the dashed black line the expected limit. The green band gives the $\pm 1 \sigma$ uncertainties of the expected limit, the yellow band the $\pm 2 \sigma$ uncertainties.}
\end{figure}
 
\begin{figure}
\centering
\begin{subfigure}[b]{0.49\textwidth}
\includegraphics[width=\textwidth]{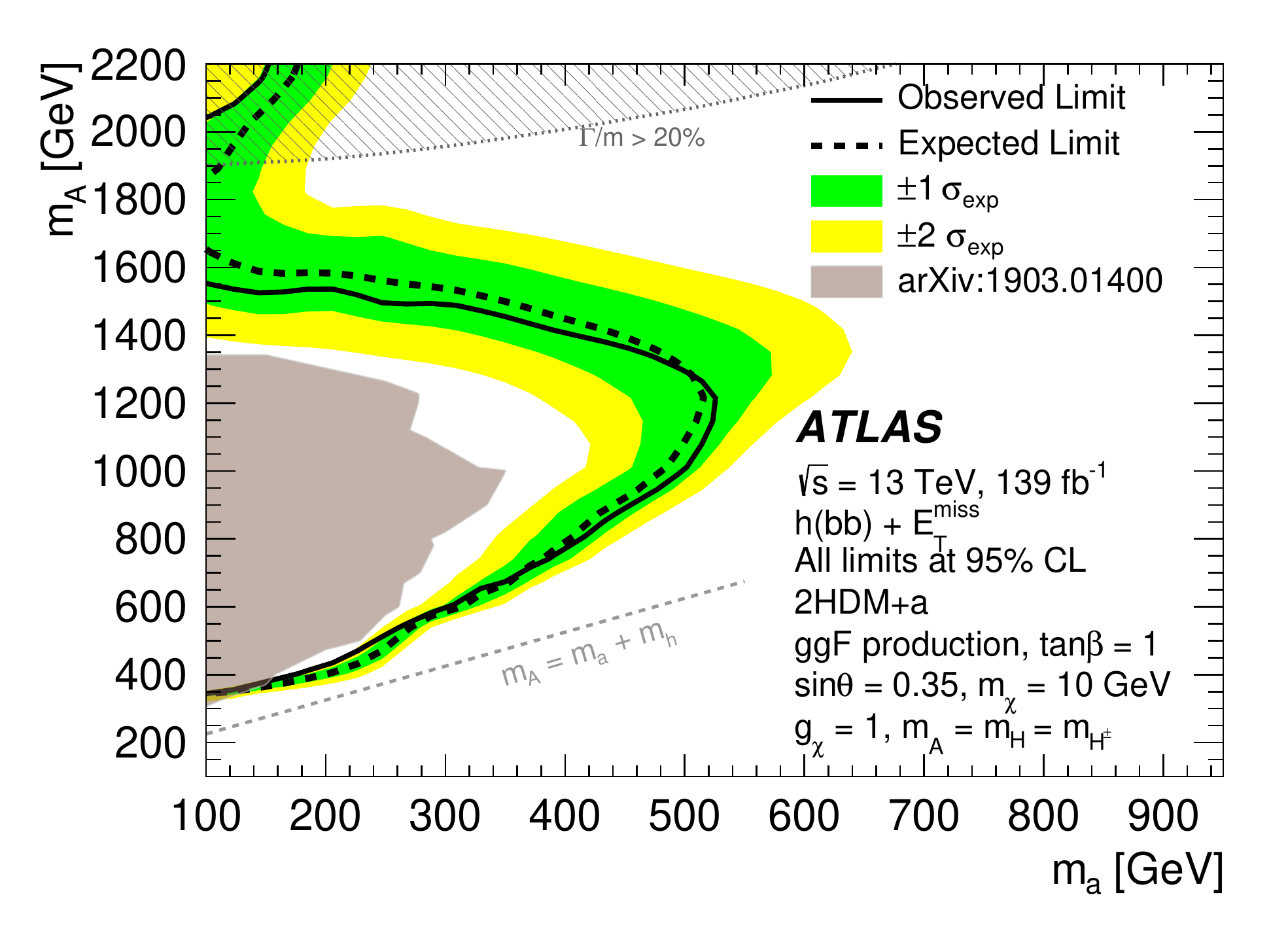}
\caption{\label{results:fig:contourplots.tb1}}
\end{subfigure}
\begin{subfigure}[b]{0.49\textwidth}
\includegraphics[width=\textwidth]{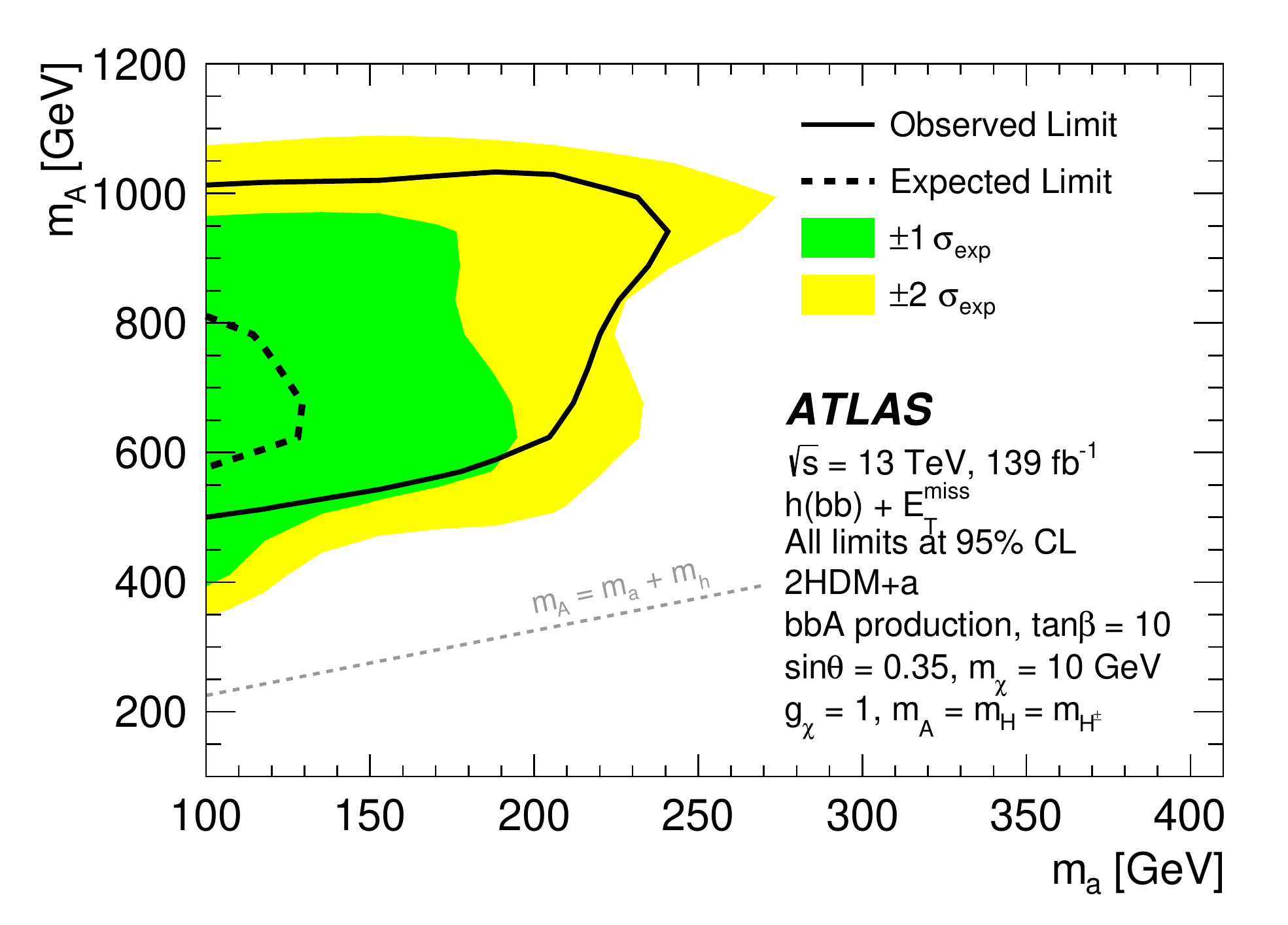}
\caption{\label{results:fig:contourplots.tb10}}
\end{subfigure}
\caption{\label{results:fig:contourplots} Exclusion limits for the 2HDM+$a$ signal with \subref{results:fig:contourplots.tb1}~$\tan \beta = 1$ and $gg$F production and with \subref{results:fig:contourplots.tb10}~$\tan \beta = 10$ and $bbA$ production. The solid black line shows the observed limit at 95\% CL, the dashed black line the expected limit. The green band gives the $\pm 1 \sigma$ uncertainties of the expected limit, the yellow band the $\pm 2 \sigma$ uncertainties. The hashed area in plot~\subref{results:fig:contourplots.tb1} indicates the region where the width of at least one of the Higgs bosons $A$, $H$ or $H^{\pm}$ or of the pseudoscalar $a$ is above 20\% of its mass.
Extending the exclusion into this region would require additional assumptions within the 2HDM+$a$ model, and as such limits are not shown beyond $2.2\,\TeV$.}
\end{figure}

Figure~\ref{results:fig:upperlimits} displays the upper limits on the visible cross-section, defined as
 
\begin{equation*}
\sigma_{\mathrm{vis},h(b\bar{b})+\mathrm{DM}} \equiv \sigma_{h+\mathrm{DM}} \times \mathscr{B}(h \to b\bar{b}) \times (\mathscr{A} \times \varepsilon)
\end{equation*}
 
where $(\mathscr{A} \times \varepsilon)$ with the acceptance $\mathscr{A}$ and the reconstruction efficiency $\varepsilon$ quantifies the probability for a certain event to be reconstructed within a window around the Higgs boson mass in a given signal region. The visible cross-section is obtained from the number of signal events in each signal region extracted from a fit to the $m(b\bar{b})$ distribution as described below, divided by the integrated luminosity.
 
In contrast to the model-specific exclusion limits in Figures~\ref{results:fig:contourplots2} and~\ref{results:fig:contourplots}, the upper limits on the visible cross-section are calculated without dependence on a signal model. They only assume that a resonance was produced with a mass close to 125~\GeV\ and decays into a pair of $b$ quarks in association with \MET. For this purpose, the binning in $m_{h}$ in the signal regions is modified such that all bins under the Higgs boson peak in a range from 90~\GeV\ to 150~\GeV\ are merged into one bin. This bin includes the Higgs boson peak, but excludes important parts of the $Z$ boson peak. A simultaneous fit to all control and signal regions is performed, using the modified binning. However, in order not to assume a specific signal model, a signal may only be present in the one bin under the Higgs boson peak. Although all signal regions are fitted simultaneously, the signal contributions in the different signal regions are independent of each other, which is ensured by using different signal strength parameters.
 
\begin{figure}
\centering
\includegraphics[width=0.7\textwidth]{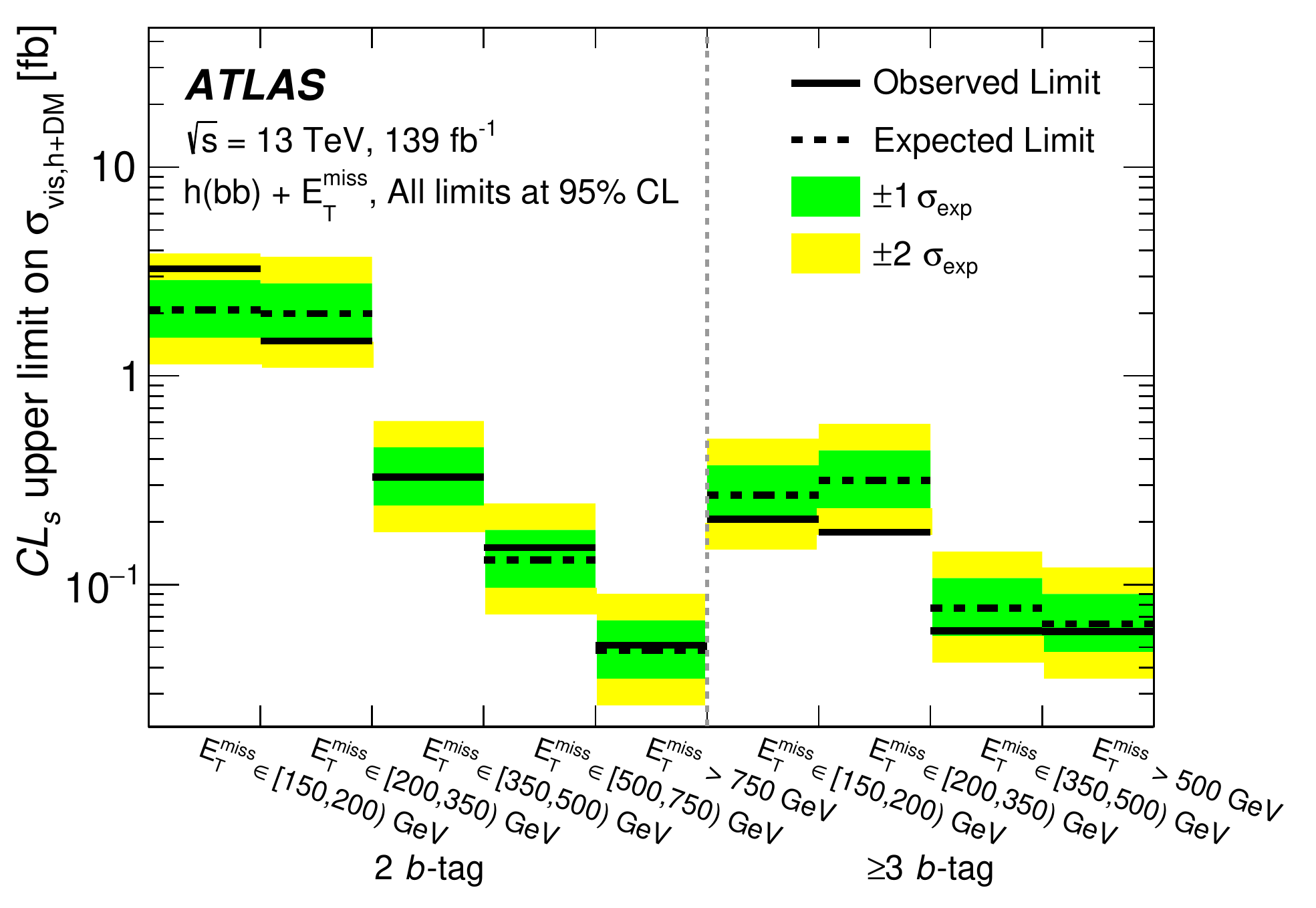}
\caption{\label{results:fig:upperlimits}Model-independent upper limits on the visible cross-section $\sigma_{\mathrm{vis},h(b\bar{b})+\mathrm{DM}} \equiv \sigma_{h+\mathrm{DM}} \times \mathscr{B}(h \to b\bar{b}) \times \mathscr{A} \times \varepsilon$ in the different signal regions.}
\end{figure}
 
 
\FloatBarrier
\section{Conclusion}
\label{sec:conclusion}
 
A search for dark matter production in association with a Higgs boson decaying into $b\bar{b}$ is presented, considering final states with either two or at least three $b$-tagged jets and without leptons. The search uses proton--proton collision data recorded by the ATLAS experiment at the LHC during the data-taking periods in 2015--2018, corresponding to an integrated luminosity of 139~\ifb. The analysis targets the 2HDM extended by dark matter particles and either a heavy vector boson $Z'$ ($Z'$-2HDM) or a pseudoscalar singlet $a$ (2HDM+$a$) as mediator particle.
 
Sensitivity to these models is obtained by considering different $b$-jet multiplicity regions, and the cases of boosted or non-boosted Higgs bosons. Further improvements relative to previous searches targeting this signature with a partial dataset from Run~2 of the LHC are obtained through better background rejection of \ttbar\ processes, better identification of $b$-hadrons from the decay of a boosted Higgs boson, and improved suppression of multijet processes by using an object-based \MET\ significance.
 
No significant deviation from Standard Model expectations was found. Exclusion limits are set on the $Z'$-2HDM and extend up to $Z'$-masses of 3~\TeV\ at 95\% CL. This is an improvement of about 700~\GeV\ relative to the analysis in Ref.~\cite{EXOT-2016-25} for heavy $Z'$ mediator masses. In the case of the 2DHM+$a$, the mass of the pseudoscalar $a$ is excluded up to 520~\GeV\ for $\tan \beta =1$ and gluon--gluon fusion production, and up to 300~\GeV\ for $\tan \beta = 10$ in $b$-associated production. Upper limits on the visible cross-section range from 0.05 to 3.26~fb, depending on the signal region.

\section*{Acknowledgements}
 

We thank CERN for the very successful operation of the LHC, as well as the
support staff from our institutions without whom ATLAS could not be
operated efficiently.
 
We acknowledge the support of ANPCyT, Argentina; YerPhI, Armenia; ARC, Australia; BMWFW and FWF, Austria; ANAS, Azerbaijan; SSTC, Belarus; CNPq and FAPESP, Brazil; NSERC, NRC and CFI, Canada; CERN; ANID, Chile; CAS, MOST and NSFC, China; COLCIENCIAS, Colombia; MSMT CR, MPO CR and VSC CR, Czech Republic; DNRF and DNSRC, Denmark; IN2P3-CNRS and CEA-DRF/IRFU, France; SRNSFG, Georgia; BMBF, HGF and MPG, Germany; GSRT, Greece; RGC and Hong Kong SAR, China; ISF and Benoziyo Center, Israel; INFN, Italy; MEXT and JSPS, Japan; CNRST, Morocco; NWO, Netherlands; RCN, Norway; MNiSW and NCN, Poland; FCT, Portugal; MNE/IFA, Romania; JINR; MES of Russia and NRC KI, Russian Federation; MESTD, Serbia; MSSR, Slovakia; ARRS and MIZ\v{S}, Slovenia; DST/NRF, South Africa; MICINN, Spain; SRC and Wallenberg Foundation, Sweden; SERI, SNSF and Cantons of Bern and Geneva, Switzerland; MOST, Taiwan; TAEK, Turkey; STFC, United Kingdom; DOE and NSF, United States of America. In addition, individual groups and members have received support from BCKDF, CANARIE, Compute Canada, CRC and IVADO, Canada; Beijing Municipal Science \& Technology Commission, China; COST, ERC, ERDF, Horizon 2020 and Marie Sk{\l}odowska-Curie Actions, European Union; Investissements d'Avenir Labex, Investissements d'Avenir Idex and ANR, France; DFG and AvH Foundation, Germany; Herakleitos, Thales and Aristeia programmes co-financed by EU-ESF and the Greek NSRF, Greece; BSF-NSF and GIF, Israel; La Caixa Banking Foundation, CERCA Programme Generalitat de Catalunya and PROMETEO and GenT Programmes Generalitat Valenciana, Spain; G\"{o}ran Gustafssons Stiftelse, Sweden; The Royal Society and Leverhulme Trust, United Kingdom.
 
The crucial computing support from all WLCG partners is acknowledged gratefully, in particular from CERN, the ATLAS Tier-1 facilities at TRIUMF (Canada), NDGF (Denmark, Norway, Sweden), CC-IN2P3 (France), KIT/GridKA (Germany), INFN-CNAF (Italy), NL-T1 (Netherlands), PIC (Spain), ASGC (Taiwan), RAL (UK) and BNL (USA), the Tier-2 facilities worldwide and large non-WLCG resource providers. Major contributors of computing resources are listed in Ref.~\cite{ATL-SOFT-PUB-2020-001}.
 

\printbibliography

\clearpage 
 
\begin{flushleft}
\hypersetup{urlcolor=black}
{\Large The ATLAS Collaboration}

\bigskip

\AtlasOrcid[0000-0002-6665-4934]{G.~Aad}$^\textrm{\scriptsize 99}$,    
\AtlasOrcid[0000-0002-5888-2734]{B.~Abbott}$^\textrm{\scriptsize 125}$,    
\AtlasOrcid[0000-0002-7248-3203]{D.C.~Abbott}$^\textrm{\scriptsize 100}$,    
\AtlasOrcid[0000-0002-2788-3822]{A.~Abed~Abud}$^\textrm{\scriptsize 34}$,    
\AtlasOrcid[0000-0002-1002-1652]{K.~Abeling}$^\textrm{\scriptsize 51}$,    
\AtlasOrcid[0000-0002-2987-4006]{D.K.~Abhayasinghe}$^\textrm{\scriptsize 91}$,    
\AtlasOrcid[0000-0002-8496-9294]{S.H.~Abidi}$^\textrm{\scriptsize 27}$,    
\AtlasOrcid[0000-0002-8279-9324]{O.S.~AbouZeid}$^\textrm{\scriptsize 38}$,    
\AtlasOrcid[0000-0001-5329-6640]{H.~Abramowicz}$^\textrm{\scriptsize 158}$,    
\AtlasOrcid[0000-0002-1599-2896]{H.~Abreu}$^\textrm{\scriptsize 157}$
% ...
\bigskip
\\

% ...
$^{27}$Physics Department, Brookhaven National Laboratory, Upton NY; United States of America.\\
% ...
$^{34}$CERN, Geneva; Switzerland.\\
% ...
$^{38}$Niels Bohr Institute, University of Copenhagen, Copenhagen; Denmark.\\
% ...
$^{51}$II. Physikalisches Institut, Georg-August-Universit\"{a}t G\"ottingen, G\"ottingen; Germany.\\
% ...
$^{91}$Department of Physics, Royal Holloway University of London, Egham; United Kingdom.\\
% ...
$^{99}$CPPM, Aix-Marseille Universit\'e, CNRS/IN2P3, Marseille; France.\\
$^{100}$Department of Physics, University of Massachusetts, Amherst MA; United States of America.\\
% ...
$^{125}$Homer L. Dodge Department of Physics and Astronomy, University of Oklahoma, Norman OK; United States of America.\\
% ...
$^{157}$Department of Physics, Technion, Israel Institute of Technology, Haifa; Israel.\\
$^{158}$Raymond and Beverly Sackler School of Physics and Astronomy, Tel Aviv University, Tel Aviv; Israel.\\
% ...

\end{flushleft}


\end{document}